\documentclass[aps,prd,twocolumn,nofootinbib,superscriptaddress]{revtex4-1}

\pdfoutput=1

\usepackage{amsmath, amssymb, amsfonts}
\usepackage{bm,bbm}
\usepackage{graphicx}
\usepackage{slashed}

\newcommand{\Secl}[1]{\label{sec:#1}}
\newcommand{\Sec}[1]{section~\ref{sec:#1}}

\newcommand{\beq}{\begin{equation}\begin{aligned}}
\newcommand{\eeq}{\end{aligned}\end{equation}}
\newcommand{\beqa}[1]{\begin{equation}\begin{alignedat}{#1}}
\newcommand{\eeqa}{\end{alignedat}\end{equation}}
\newcommand{\beqs}{\begin{subequations}\begin{eqnarray}}
\newcommand{\eeqs}{\end{eqnarray}\end{subequations}}
\newcommand{\nn}{\nonumber}

\newcommand{\bea}{\begin{eqnarray}}
\newcommand{\eea}{\end{eqnarray}}

\newcommand{\bit}{\begin{itemize}}
\newcommand{\eit}{\end{itemize}}

\newcommand{\eql}[1]{\label{eq:#1}}
\newcommand{\eq}[1]{(\ref{eq:#1})}
\newcommand{\Eq}[1]{Eq.~\eq{#1}}

\newcommand{\dd}{\mathrm{d}}

\newcommand{\fr}[2]{\dfrac{#1}{#2}}

\newcommand{\Mpl}{M_\text{P}}

\newcommand{\GeV}{\>\mathrm{GeV}}

\newcommand{\cO}{\mathcal{O}}

\newcommand{\Neff}{N_\text{eff}}
\newcommand{\Nfree}{\Neff^\text{free}}
\newcommand{\Nscatt}{\Neff^\text{scatt}}

\newcommand{\rhoDR}{\rho_\text{\tiny DR}}
\newcommand{\rhonu}{\rho_\text{{\tiny 1}$\nu$}}
\newcommand{\rhofreeDR}{\rho_\text{\tiny DR}^\text{free}}
\newcommand{\rhoscattDR}{\rho_\text{\tiny DR}^\text{scatt}}

\newcommand{\ellA}{\ell_\text{A}}

\newcommand{\ydec}{y_\text{dec}}
\newcommand{\Td}{T_\text{kd}}
\newcommand{\MP}{M_\text{P}}
\newcommand{\W}{\text{\tiny $W$}}
\newcommand{\Z}{\text{\tiny $Z$}}

\usepackage{color}
\definecolor{darkgreen}{rgb}{0,0.5,0}
\definecolor{violet}{rgb}{0.5,0.5,1}

\begin{document}
\begin{flushright}
UMD-PP-014-018\\
\end{flushright}

\title{A Hidden Dark Matter Sector, Dark Radiation, and the CMB}

\author{Zackaria Chacko}
\affiliation{Maryland Center for Fundamental Physics, Department of 
Physics, University of Maryland, College Park, MD 20742, USA}
\author{Yanou Cui}
\affiliation{Maryland Center for Fundamental Physics, Department of
Physics, University of Maryland, College Park, MD 20742, USA}
\affiliation{Perimeter Institute for Theoretical Physics, Waterloo, Ontario N2L 2Y5, Canada }
\author{Sungwoo Hong}
\affiliation{Maryland Center for Fundamental Physics, Department of
Physics, University of Maryland, College Park, MD 20742, USA}
\author{Takemichi Okui}
\affiliation{Department of Physics, Florida State University, Tallahassee, 
FL 32306, USA}

\begin{abstract}

We consider theories where dark matter is composed of a thermal relic of 
weak scale mass, whose couplings to the Standard Model (SM) are however 
too small to give rise to the observed abundance. Instead, the abundance 
is set by annihilation to light hidden sector states that carry no charges 
under the SM gauge interactions. In such a scenario the constraints from 
direct and indirect detection, and from collider searches for dark 
matter, can easily be satisfied. The masses of such light hidden states can be
protected by symmetry if they are Nambu-Goldstone bosons, fermions, or gauge bosons. These states can 
then contribute to the cosmic energy density as dark radiation, leading to 
observable signals in the cosmic microwave background (CMB\@). 
Furthermore, depending on whether or not the light hidden sector states 
self-interact, the fraction of the total energy density that 
free-streams is either decreased or increased, leading to characteristic 
effects on both the scalar and tensor components of the CMB anisotropy 
that allows these two cases to be distinguished. The magnitude of these 
signals depends on the number of light degrees of freedom in the hidden 
sector, and on the temperature at which it kinetically decouples from 
the SM\@. We consider a simple model that realizes this scenario, based on 
a framework in which the SM and hidden sector are initially in thermal 
equilibrium through the Higgs portal, and show that the resulting 
signals are compatible with recent Planck results, while large enough to 
be detected in upcoming experiments such as CMBPol and CMB Stage-IV.
Invisible decays of the Higgs into hidden sector states at colliders can
offer a complementary probe of this model. 

\end{abstract}


\maketitle

\section{Introduction}

\Secl{intro}

In the last two decades, with the advent of precision cosmology, it has 
become clear that some form of non-luminous dark matter (DM) contributes 
more than 20\% of the total energy density of the 
universe~\cite{Planck:2015xua}. Although it is known that the particles of 
which DM is composed lie outside the SM of particle 
physics, their precise nature remains to be understood.

In the absence of a detailed understanding about the properties of dark 
matter, many different candidates have been put forward. A large class 
of well-motivated theories are based on the `Weakly Interacting Massive 
Particle' (WIMP) paradigm. In its simplest incarnation, this scenario 
involves a particle of weak scale mass, the WIMP, that has interactions 
of weak scale strength with the SM fields. This class 
of theories possesses the very attractive feature that the WIMPs that 
survive after their annihilation into SM particles freezes out naturally 
tend to have a relic abundance that is in good agreement with 
observations~\cite{Lee:1977ua, Vysotsky:1977pe}.

In this conventional scenario, the WIMP must have interactions of weak 
scale strength with the SM fields. Several different types of DM 
experiments are searching for evidence of such interactions. These 
include direct detection experiments that are looking for the recoils of 
nuclei after being impacted by a DM particle, indirect detection 
experiments that seek to observe the products of DM annihilation, and 
collider experiments such as the the Large Hadron Collider (LHC) that seek to produce DM\@. Till date, 
there has been no compelling evidence for the existence of such 
interactions, and the experimental limits now exclude a significant part 
of the preferred parameter space for many WIMP DM 
candidates~\cite{Aprile:2012nq, Akerib:2013tjd, Ackermann:2013yva, 
Abdo:2010ex, Djouadi:2011aa}.

With the simplest realizations of the WIMP paradigm beginning to come 
under strain, several ideas have been put forward to explain the absence 
of a signal in these experiments. Among the hypotheses that have been 
advanced are that the DM candidate scatters 
inelastically~\cite{TuckerSmith:2001hy, Cui:2009xq}, is 
leptophilic~\cite{Krauss:2002px, Baltz:2002we, Kopp:2009et}, or 
interacts preferentially with heavier quark flavors~\cite{Cheung:2010zf, 
Agrawal:2011ze, Zhang:2012da, Kumar:2013hfa, Kilic:2015vka}. An 
alternative proposal that has attracted 
interest~\cite{Finkbeiner:2007kk,Pospelov:2007mp,Feng:2008ya,Feng:2008mu} 
is the idea that, while DM is indeed composed of WIMPs, their couplings 
to the SM fields are suppressed, and too small to yield the observed 
abundance. Instead, the DM candidate possesses interactions of weak 
scale strength with a new hidden sector that carries no charge under the 
SM gauge interactions, and its relic abundance is set by its 
annihilation into these states. Such a scenario can naturally account 
for the observed abundance of DM, while explaining the absence of any 
signal in experiments.

It is not difficult to envisage scenarios where the DM candidate 
naturally has weak scale mass and interactions of weak scale strength 
with a hidden sector. For example, in supersymmetric theories, both the 
weak scale and the scales in the hidden sector could be set by the scale 
of supersymmetry breaking. Similarly, in extra dimensional 
Randall-Sundrum constructions, both the Higgs and the hidden sector 
states could be localized to the infrared brane. In such a scenario the 
scales in the hidden sector would again naturally be of order the weak 
scale. Therefore, provided the SM and hidden sectors are in thermal 
equilibrium at or above the weak scale, so that their temperatures at 
freeze out are not very different, this framework can naturally explain 
the observed abundance of DM while remaining consistent with all 
experimental constraints.

The existence of a hidden sector into which DM annihilates can 
potentially be tested by experiments. The nature of the signals depends 
on the masses of the particles in the hidden sector, and on their 
couplings to SM states. If all the particles in the hidden sector have 
masses above an eV, and the temperature of this sector is comparable 
that of the SM, we expect these states will decay or annihilate into SM 
particles before the CMB epoch. This is because the lightest state in 
the hidden sector, being massive, would otherwise contribute to the 
energy density in DM, violating the overclosure bounds if it is heavier 
than a keV, and coming into conflict with the cosmological constraints 
on a warm sub-component of DM if it is lighter than a keV. Such a 
scenario therefore implies the existence of couplings between the hidden 
sector states and the SM that can potentially be tested in experiments, 
as in the scenarios of exciting 
DM~\cite{Finkbeiner:2007kk,Cholis:2008vb}, secluded 
DM~\cite{Pospelov:2007mp} and boosted DM~\cite{Agashe:2014yua, 
Berger:2014sqa}. If instead, some or all of the states in the hidden 
sector have masses below an eV, they would be expected to constitute a 
significant component of the energy density of the universe both before 
and during the epoch of matter-radiation equality, potentially leading 
to observable signals in the 
CMB~\cite{Feng:2008ya,Feng:2008mu,Ackerman:mha,Feng:2009mn,Franca:2013zxa, 
Weinberg:2013kea,Garcia-Cely:2013nin,Garcia-Cely:2013wda}. The simplest 
possibility is that these states, if present, are massless, and 
constitute dark radiation (DR) at present times, thereby obviating the 
need for any other mass scales in the theory. It is this scenario, and 
the associated signals, that we will focus on in this 
paper.{\footnote{If the DM and DR are tightly coupled, oscillations of 
the DM-DR fluid can also give rise to signals in the matter power 
spectrum~\cite{Blennow:2012de,Diamanti:2012tg}, (see 
also~\cite{Foot:2014uba,Foot:2014osa}). However, the large strength of 
the interaction required to obtain an observable effect would overly 
deplete the abundance of DM, and is therefore disfavored in the 
framework of thermal WIMPs.}}

The presence of these new light particles implies the existence of 
additional structure in the theory, if the scenario is to be natural. 
There are three known symmetries that can prevent masses from being 
generated for a massless particle: a shift symmetry for a spin-0 
Nambu-Goldstone boson, a chiral symmetry for a spin-$1/2$ fermion, and a 
gauge symmetry for a spin-$1$ vector boson. DR candidates protected by 
these symmetries have been considered, for example, 
in~\cite{Nakayama:2010vs, Fischler:2010xz}. These symmetries may appear 
individually or in combination; for example, the spectrum of light 
states may consist of a single Nambu-Goldstone boson, but it may also 
consist of spin-$1/2$ fermions charged under a U(1) gauge group with its 
associated massless spin-1 boson~\cite{Jeong:2013eza}. This latter 
example illustrates that the constituents of the DR need not be free, 
but may have interactions amongst themselves without violating the 
symmetries that keep them light. In general, therefore, we see that the 
DR can take two distinct forms:
 \begin{itemize}
 \item \emph{Free DR,} which free streams during the era of acoustic 
oscillations, and is characterized by a mean free path $\gg H^{-1}$, 
where $H$ is the Hubble constant.
 \item \emph{Scattering DR}, which scatters during the era of acoustic
oscillations, and is characterized by a mean free path $\ll H^{-1}$.
 \end{itemize} 
 Since the presence of DR is a robust prediction of this scenario, it is 
important to understand whether it can be detected, and whether we can 
distinguish between the two different cases of free DR and scattering 
DR\@. It is these questions that we shall be primarily concerned with in 
this paper. We find that, provided the hidden DM sector was in thermal 
equilibrium with the SM at temperatures at or above the weak scale, the 
contribution of the DR to the energy density during the CMB epoch is in 
general large enough to be detected in future experiments, such as 
CMBPol \cite{Galli:2010it} ($\sigma_{N_{\rm eff}}=0.044$), and 
eventually CMB Stage-IV \cite{Abazajian:2013oma} ($\sigma_{N_{\rm 
eff}}=0.02$).

We also find that it is, in general, possible to distinguish between 
scenarios with free streaming DR and scattering DR\@. Studies of the 
scalar~\cite{Peebles, Hu:1995en, Bashinsky:2003tk} and 
tensor~\cite{Weinberg:2003ur, Flauger:2007es} metric perturbations have 
established that the details of the CMB spectrum depend on the fraction 
of the energy density in radiation that is free streaming. This ratio 
impacts not just the amplitudes of the modes, but also the locations of 
the peaks in the CMB spectrum. In scenarios where neutrinos scatter off 
new light states during the period immediately prior to the CMB epoch, 
as in models of late time neutrino masses~\cite{Chacko:2003dt, 
Chacko:2004cz,Okui:2004xn}, and in the neutrinoless universe 
scenario~\cite{Beacom:2004yd}, this ratio differs significantly from the 
SM prediction. Consequently, it has been possible to establish that this 
class of theories, which was already disfavored by the WMAP 
data~\cite{Hannestad:2004qu, Trotta:2004ty, Bell:2005dr, Cirelli:2006kt, 
Friedland:2007vv}, is now excluded by Planck, unless the new neutrino 
interactions come into equilibrium only very shortly prior to 
matter-radiation 
equality~\cite{Archidiacono:2013dua,Forastieri:2015paa}.

Similar considerations apply to the class of theories we are 
considering. In the presence of a new dark component of radiation, the 
free streaming fraction is altered, with the sign of the correction 
dependent on whether the DR scatters or free streams. Consequently, the 
amplitudes of the scalar and tensor modes receive corrections, with the 
sign of the effect dependent on whether the DR is free or 
self-interacting. In addition, the locations of the CMB peaks are 
shifted, with the sign of the shift again dependent on whether or not 
the DR carries self-interactions. We find that these effects may be 
large enough to allow upcoming experiments to distinguish between free 
DR and scattering DR\@. Therefore the CMB offers a window into the 
dynamics of the hidden sector that DM annihilates into.

The outline of this paper is as follows. In the next section we discuss 
the CMB signals associated with DR, and explain how scenarios with free 
streaming DR and scattering DR can be distinguished. In section III we 
show how the expression for the relic abundance of DM may be generalized 
to the case when the temperature of the dark sector differs from that of 
the SM\@. In section IV we consider a simple model based on the Higgs 
portal that realizes this scenario, and show that the signals can be 
large enough to be detected by upcoming CMB experiments, while remaining 
compatible with the recent Planck results~\cite{Planck:2015xua}. We also 
show that invisible decays of the Higgs into hidden sector states at 
colliders can offer a complementary probe of this scenario. Our 
conclusions are in section V.

\section{The CMB Signals of Dark Radiation}
\Secl{CMBsignals}

The CMB spectrum is affected by the presence of extra relativistic 
degrees of freedom during the era between matter-radiation equality and 
photon decoupling. It is customary in cosmology to quantify the 
contribution to the energy density from such additional radiation in 
unit of the energy density of a single relativistic SM neutrino species 
$\rhonu$,
 \beq
\Delta \Neff 
\equiv 
\fr{\rhoDR}{\rhonu} 
\,,\eql{def:Neff}
 \eeq
 where $\rhoDR$ is the energy density of DR, and all SM neutrinos are 
treated as being relativistic at the temperatures in question.

For any specific hidden sector model, we can calculate $\rhoDR$, and 
hence $\Delta \Neff$. The first step is to determine the temperature of 
the dark sector, $\hat{T}$, that corresponds to a given SM temperature 
$T$ at the same cosmic time $t$. To do this, note that the comoving entropies of the SM and of the 
dark sector are separately conserved after the two sectors thermally 
decouple from each other. Then, after  thermal 
decoupling when $T= \hat{T}= \Td$, but before the neutrino 
decoupling, taking the ratio of the two entropy conservation 
relations leads to the relation
 \begin{equation}
\frac{\hat{g}_* \, \hat{T}^3}{g_* \, T^3}
 =
\frac{\hat{g}_{*\text{kd}}}{g_{*\text{kd}}}
 \,.
 \eql{temp_ratio}
 \end{equation}
 Here $g_*$ and $g_{* \rm kd}$ are the number of degrees of freedom in 
the SM at temperatures $T$ and $\Td$ respectively, with the usual $7/8$ 
factors for the fermions. The corresponding parameters in the dark 
sector are labelled by $\hat{g}_*$ and $\hat{g}_{* \rm kd}$. Applying 
this relation just above the neutrino decoupling temperature $T \sim 
\mathcal{O}(10)$~MeV, $g_{*} = 10.75$, and noting that the contribution 
of a single neutrino species to the energy density is given by $\rhonu = 
\frac{7}{4}\frac{\pi^2}{30} T^4$, we have
 \begin{equation}
\Delta \Neff
=
\frac{\hat{g}_* \, \hat{T}^4}{\frac74 \, T^4}
=
\frac{4}{7} \hat{g}_*
\!\left( \frac{{g}_{*}}{\hat{g}_*}
         \frac{\hat{g}_{*\text{kd}}}{g_{*\text{kd}}}
\right)^{\!\! 4/3}.
 \label{DeltaN}
 \end{equation}
 Note that the above $\Delta \Neff$ computed for the time just before 
neutrino decoupling is the same as the $\Delta \Neff$ at the later CMB 
time,
 as $\hat{T}$ and $T_\nu$ redshifts the same way till then.
 
As outlined in \Sec{intro}, in general $\rhoDR$ can consist of 
two qualitatively very different types of radiation: free-streaming 
radiation with mean free path $\gg H^{-1}$, and scattering radiation 
with mean free path $\ll H^{-1}$. We can parametrize each of these 
components of radiation in complete analogy with the 
definition~\eq{def:Neff}, 
 \beq
\Delta \Nfree 
\equiv 
\fr{\rhofreeDR}{\rhonu} 
\,,\quad
\Delta \Nscatt 
\equiv 
\fr{\rhoscattDR}{\rhonu}
\,,
 \eeq
 so that the total extra radiation $\Delta \Neff = \Delta \Nfree + 
\Delta \Nscatt$. 

\begin{figure}[t]
 \includegraphics[height=57mm]{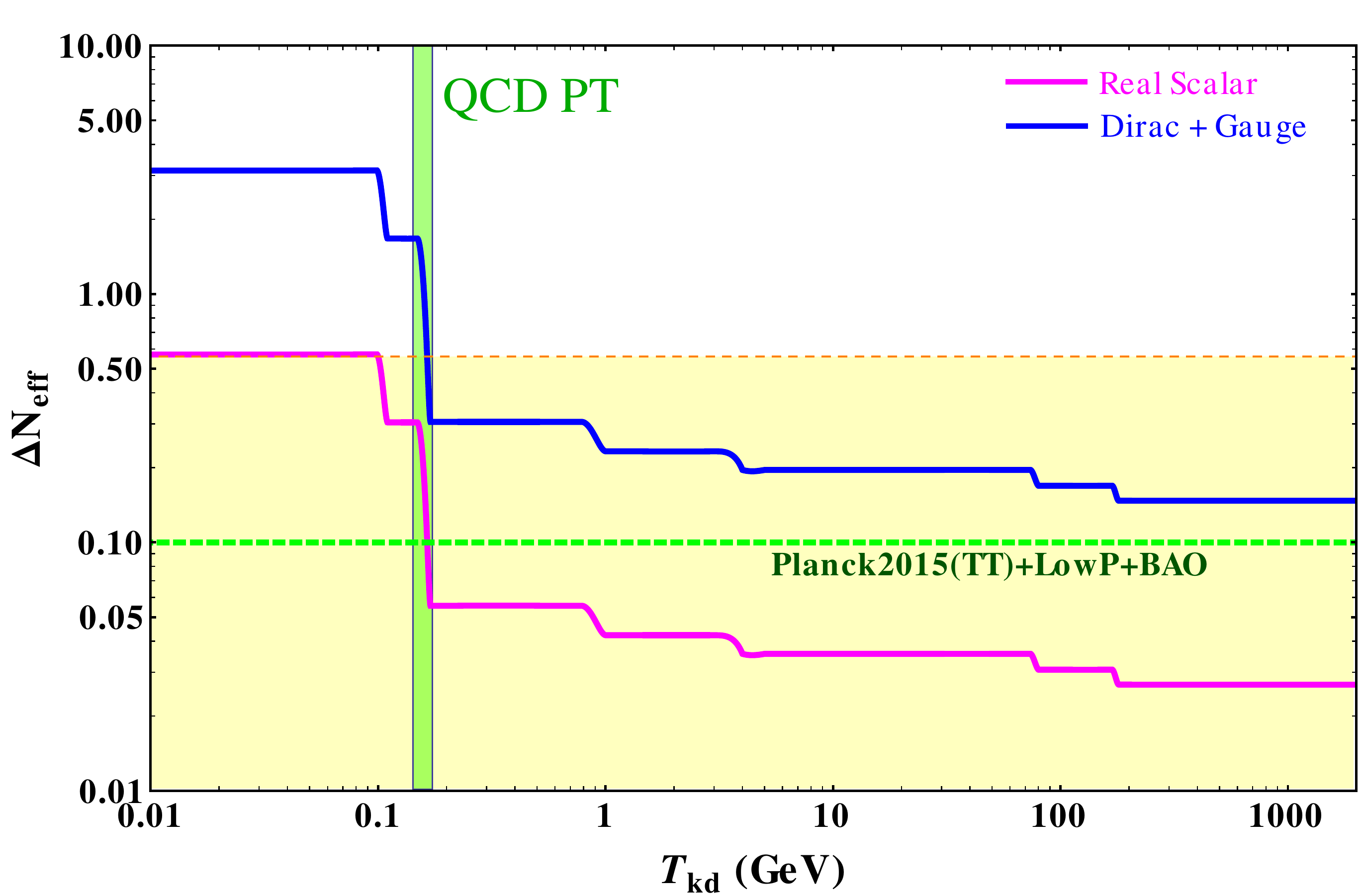} 
  \caption{ $\Delta N_{\text{eff}}$ as a function of the temperature at which the SM and dark sector thermally decouple. Also shown are the 2015 Planck results: the central value (Green dashed line) and the $2 \sigma$ constraint (Orange dashed line).}
\label{fig:N_eff}
\end{figure}

In this class of models, for a given thermal decoupling temperature 
$\Td$, there is a robust lower bound on $\Delta \Neff$. To understand 
this, note that the lowest possible value of $\hat{g}_*$ in 
Eq.~(\ref{DeltaN}) is 1, corresponding to the case when the dark 
radiation consists of just a single real scalar. Then, if thermal 
decoupling between the hidden sector and the SM occurs at temperatures 
well below the mass of the DM particle, we can have $\hat{g}_{* \rm kd} = 
\hat{g}_* = 1$. In this limit we obtain a lower bound on $\Delta \Neff$,
  \begin{equation}
\Delta \Neff^{\rm min}
=
\frac{4}{7}\!\left( \frac{{g}_{*}}{g_{*\text{kd}}}\right)^{\!\! 4/3}.
 \label{DeltaNmin}
 \end{equation}
 From Eq.~(\ref{DeltaNmin}), assuming all the SM degrees of freedom are 
in the bath at decoupling, we have $g_{*\rm kd}=106.75$, which leads to 
a lower bound on the effective number of neutrinos, $\Delta \Neff^{\rm 
min}\gtrsim 0.027$. This result applies to arbitrarily high $\Td$ 
provided there are no new states in the SM sector up to that scale. In 
Fig.~\ref{fig:N_eff} we have plotted this lower bound as a function of 
the decoupling temperature $\Td$. Quite intriguingly, the ultimate 
experimental sensitivity at CMB-Stage-IV is $\sigma_{ \Neff}=0.02$ 
\cite{Abazajian:2013oma}, which would allow it to probe the DM scenario 
we outline here. This projected experimental sensitivity is based on a 
one parameter extension of the standard six parameter $\Lambda$CDM model 
that accommodates varying $\Neff$. Although this projection assumes that 
the primordial density fluctuations have an exactly power law spectrum, 
the recent Planck results have established that at present this is an 
excellent fit to data~\cite{Planck:2015xua}, and so this forms a 
reasonable basis for estimating the sensitivity.

If the dark sector consists of just a real scalar, we expect that it is 
a Goldstone boson, so that its mass is protected against radiative 
corrections from the weak scale. In this scenario, the interactions of 
the states that constitute the DR are momentum suppressed, and so the DR 
free streams. In scenarios where the DR possesses self interactions 
large enough to prevent free streaming, the requirement of naturalness 
up to the weak scale implies that it must be composed of more than just 
a single real scalar, or else the radiative corrections to the scalar 
mass from the self interactions would tend to make its mass much greater 
than an eV. Therefore, in scenarios where the DR scatters, we expect 
that there will be additional light states in the hidden sector, and so 
$\Delta \Neff$ is expected to be larger than its minimum value, $\Delta 
\Neff^{\rm min}$. In Fig.~\ref{fig:N_eff}, we have plotted $\Delta 
\Neff$ as a function of $\Td$ for the case when the DR consists of a 
pair of massless Weyl fermions with vector-like charges under a U(1) 
gauge group, and the associated massless gauge boson. We see that even 
for high $\Td$, we predict $\Delta \Neff \gtrsim 0.15$, which is large 
enough to be observed at CMBPol.

In the rest of this section we discuss how the experimental limits on 
$\Delta \Neff$ are obtained, and how we can distinguish between the two 
cases of scattering DR and free streaming DR\@.

\subsection{The Determination of $\Delta \Neff$} 

At present, limits on $\Delta \Neff$ are obtained by considering how 
the presence of additional energy density in radiation would affect the 
quality of the fit in the six parameter $\Lambda$CDM model. Of the six 
parameters, two are particularly sensitive to $\Delta \Neff$. These are 
the total energy density in matter, $\rho_m$, which is the sum of the 
energy densities in baryons and DM, and in the cosmological constant, 
$\rho_\Lambda$. The presence of additional energy density in radiation 
would tend to delay the onset of matter-radiation equality. Since the 
amplitude of a Fourier mode is very sensitive to the fraction of energy 
density in matter as it crosses the horizon, this ratio is highly 
constrained by the CMB data. Therefore, for $\Delta \Neff > 0$, the best 
fit is obtained by increasing $\rho_m$ in the appropriate proportion to 
ensure that the redshift at the onset of matter-radiation equality is 
unaffected. Since the energy density in baryons, $\rho_b$, is very 
tightly constrained by measurements of the relative heights of the even 
and odd CMB peaks and cannot be altered, the change in $\rho_m$ is 
accomplished by an increase in the energy density in DM\@.

The additional energy density in DR, and in matter, then implies an 
increase in the Hubble constant during the CMB epoch. This will in turn 
affect the size of the sound horizon, leading to a change in the 
locations of the CMB peaks. This observable is, once again, highly 
constrained by the data. However, this effect can be offset by changing 
$\rho_\Lambda$ so as to alter the distance to the last scattering 
surface, thereby keeping the angular locations of the peaks intact. 
Nevertheless, as we now explain, the change in the Hubble constant 
during the era of acoustic oscillations leads to other effects in the 
CMB spectrum that can no longer be compensated for once $\rho_m$ and 
$\rho_\Lambda$ are fixed.

Prior to recombination, the photons interacted strongly with the 
baryons. Although the photon mean free path during this era was 
relatively short, the photons were nevertheless able to diffuse outward, 
with a characteristic diffusion length $r_d$. As a consequence of this 
diffusion, inhomogeneities and anisotropies at scale smaller than the 
$r_d$ are suppressed. This damps the peak amplitudes at higher $\ell$ 
relative to the first peak at $\ell\simeq220$, which corresponds to 
modes that entered the horizon near recombination. This effect is known 
as Silk damping, or diffusion damping. A change in the Hubble rate 
affects the time available for diffusion, leading to observable effects 
on the CMB spectrum. In particular, the height of the first CMB acoustic 
peak relative to the latter peaks is altered. Therefore, this effect can 
be used to place limits on the Hubble constant during the epoch of 
acoustic oscillations, and therefore on $\Delta \Neff$.  The
presence of additional energy density in radiation also leads to changes
in the CMB spectrum associated with the early Integrated Sachs-Wolfe
(ISW) effect, but these are less significant than the effects arising
from Silk damping~\cite{Hou:2011ec}.

In principle, an increase in the fraction of baryons in helium, $Y_{\rm 
He}$, while $\rho_b$ is held fixed, would reduce the number of free 
electrons available for scattering, and could also account for a change 
in the scale of Silk damping. However, the helium fraction in the SM can 
be calculated sufficiently precisely from Big Bang nucleosynthesis so as 
to exclude this as the explanation for any observed discrepancy. For a 
good discussion of these issues with more details, see~\cite{Hou:2011ec, 
Brust:2013ova}.

\subsection{Distinguishing between Free and Scattering DR 
via Scalar Metric Perturbations}

Several authors have considered the effects of the SM neutrinos on the 
scalar component of the CMB spectrum~\cite{Peebles, Hu:1995en, 
Bashinsky:2003tk}. These results can easily be generalized to the case 
when there is additional energy density in radiation, and can be used to 
distinguish between free streaming DR and scattering DR\@.

In the conformal Newtonian gauge the Robertson-Walker metric with 
scalar perturbations takes the form,
 \begin{equation}
\mathrm{d} s^2 = a^2 (\tau) \left( - \left( 1 + 2 \Phi \right) \mathrm{d} \tau^2 + \left( 1- 2 \Psi \right) \mathrm{d} \mathbf{r}^2  \right)
 \end{equation}
 Here $\tau$ represents conformal time, while $a$ is the cosmological 
scale factor. $\Psi$ and $\Phi$ represent the scalar metric 
perturbations. In the absence of any free streaming particle species, we 
have $\Psi = \Phi$. When, however, a free streaming species is present, 
the energy momentum tensor becomes anisotropic. This leads to a 
difference between $\Psi$ and $\Phi$ that is proportional to $f_\nu$, 
the total energy density in free streaming radiation expressed as a 
fraction of the total energy density in radiation.
 \beq
f_\nu
\equiv
\fr{\rho_\text{all free rad}}{\rho_\text{all rad}}
=
\fr{3\rhonu + \rhofreeDR}{3\rhonu + \rho_\gamma + \rhofreeDR + \rhoscattDR}
\,.\eql{fnu}
 \eeq
In the limit that $\rhofreeDR$ and $\rhoscattDR$ are small compared to 
$\rho_\text{all rad}$, the total energy density in radiation, 
the deviation from the standard cosmology is given by
\begin{align}
& f_\nu - f_\nu\bigr|_\text{SM}
\nonumber\\
&= \frac{f_\nu\bigr|_\text{SM}}{3} \Bigl[ \!\left( 1 - f_\nu\bigr|_\text{SM} \right) \Delta\Nfree - f_\nu\bigr|_\text{SM} \, \Delta\Nscatt \Bigr]
\nonumber\\
&= \frac{0.41}{3} \left(0.59 \Delta \Nfree - 0.41 \Delta \Nscatt \right) .
\eql{Delta_fnu}
\end{align}

Now, the solution of the coupled system of equations for matter, 
radiation and gravity reveals that the presence of a free streaming 
component in radiation is associated with a change in the amplitudes of 
the CMB modes at large $\ell$. The magnitude of this effect was first 
determined numerically in~\cite{Peebles}. Subsequently, analytic 
expressions were obtained in~\cite{Hu:1995en,Bashinsky:2003tk}. The 
result is given by,
 \begin{equation}
\frac{\delta C_\ell}{C_\ell} = - \frac{8}{15} f_\nu \; .
 \end{equation} Then, using \Eq{Delta_fnu}, we can obtain an expression 
for the fractional change in $C_\ell$ with respect to the standard 
cosmology,
  \begin{eqnarray}
\frac{\Delta C_\ell}{C_\ell} 
&=& \frac{\delta C_\ell}{C_\ell} - \frac{\delta C_\ell}{C_\ell}\biggl|_\text{SM} 
\nonumber\\
&=& -\frac{8}{15} \!\left( f_\nu - f_\nu\bigr|_\text{SM} \right)
\nonumber \\
&=& -0.072 \left(0.59 \Delta \Nfree - 0.41 \Delta \Nscatt \right) .
 \end{eqnarray}
 We see that the result is independent of $\ell$, and that the sign of 
this effect depends on whether the DR is scattering or free streaming.

 In addition to the corrections to the amplitude, there is a shift in 
the angular locations of the high $\ell$ CMB peaks by an equal 
amount~\cite{Bashinsky:2003tk}. This signal is particularly important 
because, in contrast to other effects of DR such as Silk damping, it is 
difficult to mimic by altering other parameters such as the helium 
fraction. The magnitude of this shift is again proportional to the free 
streaming fraction $f_\nu$,
 \beq
\delta\ell 
\>\>\simeq\> 
-57 \, f_\nu \, \fr{\ellA}{300} \,.
\eql{BS}
 \eeq
 Here $\ellA\approx 300$ represents the average angular spacing between 
the CMB peaks at large $\ell$. Again, in the limit that DR contributes 
only a small fraction of the total energy in radiation, $\Delta \Neff 
\lesssim 1$, \Eq{Delta_fnu} leads to an expression for the change in $\delta\ell$ 
with respect to the standard cosmology,
 \begin{align}
\Delta\ell &\equiv \delta\ell - \delta\ell\bigl|_\text{SM} 
\nn\\
&=
-57 \!\left( f_\nu - f_\nu\bigr|_\text{SM} \right)\! \fr{\ellA}{300} 
\nonumber \\
&\simeq
-7.8 
\left( 
0.59 \Delta \Nfree - 0.41 \Delta \Nscatt
\right) \!
\fr{\ellA}{300}
\,.\eql{Delta_ell}
 \end{align}
 Once again we see that the sign of the effect depends on whether the DR 
is scattering or free streaming.

We can obtain a very rough estimate of the sensitivity of upcoming CMB 
experiments to the effects of $\Delta \ell$ by considering how well 
$\Neff$ can be determined when the helium fraction $Y_{\rm He}$ is 
allowed to float freely. In this limit, the effects of $\Delta \Neff$ on 
Silk damping can be compensated for by changes in $Y_{\rm He}$. Under 
these circumstances, the shifts in the locations of the CMB peaks play 
an important role in the determination of $N_{\rm eff}$, and we can 
interpret the results as a rough guide to the sensitivity of these 
experiments to $\Delta \Neff$ arising from its effect on $\Delta \ell$, 
and not its effect on Silk damping. The projected sensitivity of CMBPol 
to $\Neff$ when $Y_{\rm He}$ is allowed to float is $\Delta \Neff = 
0.09$~\cite{Bashinsky:2003tk}. We therefore expect that provided $\Delta 
\Neff \gtrsim 0.10$, upcoming experiments will have some sensitivity to 
whether DR is free streaming or scattering, allowing the possibility of 
distinguishing between these two scenarios.

\subsection{Distinguishing between Free and Scattering DR 
via Tensor Metric Perturbations}

The presence of a free streaming component of radiation also affects the 
tensor component of the CMB spectrum. Detailed studies of the effects of 
the SM neutrinos on the tensor modes (the $B$- and $E$-modes) of the CMB 
were performed in~\cite{Weinberg:2003ur, Flauger:2007es}, which found an 
$\cO(10)$\% damping of the correlation functions of the tensor modes at 
long wavelengths, rising to an $\cO(35)$\% damping at short wavelengths. 
Analytic results for the damping were subsequently obtained 
in~\cite{Dicus:2005rh, Stefanek:2012hj}. The corrections to the spectrum 
that arise from the presence of a free streaming DR component were 
considered in~\cite{Jinno:2012xb}.
 
We now show that the results of~\cite{Weinberg:2003ur} can be 
generalized in a very simple way to arbitrary $\Neff$, provided $\Delta 
\Neff \lesssim 1$. The crucial observation is that, in the analysis 
of~\cite{Weinberg:2003ur}, the effects of the SM neutrinos arise 
entirely from their contribution to $\bar{f}_\nu$, the free-streaming 
fraction of the total energy density,
 \beq
\bar{f}_\nu
\equiv
\fr{\rho_\text{all free rad}}{\rho_\text{total}}
=
\fr{3\rhonu + \rhofreeDR}{\rho_\text{total}}
\,.\eql{barfnu}
 \eeq 
 Therefore, by understanding how the result depends on $\bar{f}_\nu$, we 
can immediately determine how the correlation functions of the tensor 
modes depend on $\Delta \Nfree$ and $\Delta \Nscatt$. During the 
radiation dominated era, to a very good approximation, $\bar{f}_\nu = 
f_\nu$. However, as matter-radiation equality approaches, the 
contribution of matter to the total energy density can no longer be 
neglected, and $\bar{f}_\nu$ and $f_\nu$ are distinct. 

The Robertson-Walker metric with tensor perturbations takes the form,
 \begin{equation}
\mathrm{d} s^2 = a^2 (\tau) \left( - \mathrm{d} \tau^2 + \left[ \delta_{ij} + h_{ij} (\mathbf{x}, \tau) \right] \mathrm{d} x^i \mathrm{d} x^j \right)
 \end{equation}
 with $h_{ii} = \partial_i h_{ij} = 0$. We define a new coordinate $u = 
k \tau$, where $k$ is the co-moving wave number. Then, as shown 
in~\cite{Weinberg:2003ur}, the amplitude of tensor perturbations with 
co-moving wave number $k$ can be written as $h_{ij}(u) = h_{ij}(0) \, 
\chi(u)$, where the function $\chi(u)$ remains to be determined. It
satisfies the integro-differential equation
 \beq
{\mathcal{F}(u, \partial_u)} \, \chi(u) = \bar{f}_\nu \int_0^u \mathcal{I}(u,U) \, \chi'(U) \, \dd U
\,,
\label{weinberg_eqn}
 \eeq
 where $\mathcal{F}(u, \partial_u)$ is a second-order, linear 
differential operator. Its precise form, along with that of the kernel 
function $\mathcal{I}(u, U)$, may be found in~\cite{Weinberg:2003ur}. 
The initial conditions on $\chi$ are given by $\chi(0) = 1$ and 
$\chi'(0) = 0$.

In general, the integro-differential equation Eq.~(\ref{weinberg_eqn}) is 
not simple to solve. However, approximate analytic solutions that apply 
in certain limits have been obtained for the important case of the SM 
with 3 free streaming neutrinos, $\Neff = 3.046$. As we now explain, 
these solutions can be generalized in a simple way to the case of 
arbitrary $\bar{f}_\nu$, provided $\Delta \Neff \lesssim 1$.

For \textit{short wavelengths} that enter the horizon well before matter 
radiation equality, that is, $u\gg1$, the solution to 
Eq.~(\ref{weinberg_eqn}) approaches a homogeneous 
solution~\cite{Weinberg:2003ur},
 \beq
\chi(u)\to A \, \fr{\sin(u+\delta)}{u} \,,
 \eeq
 where the parameters $A$ and $\delta$ contain the dependence on 
$\bar{f}_\nu = f_\nu$. In the limit $f_\nu=0$, $A$ and $\delta$ take 
values $A_0=1, ~\delta_0=0$. We denote this solution by $\chi_0(u)$.
For the case of the three free streaming 
neutrino species of the SM, with $f_\nu = f_\nu^\text{SM} = 0.41$, a 
numerical study~\cite{Weinberg:2003ur} leads to the values $A_{\rm SM} = 0.80,~\delta_{\rm 
SM}\approx 0$. The fact that the value of $A$ changes by only about 
$20\%$ for the change from $\Neff = 0$ to $\Neff = 3.046$ indicates that for 
$\Delta \Neff \lesssim 1$ the term proportional to $f_\nu$ can be 
treated as a perturbation. Accordingly, we may obtain an approximate 
solution for general $f_\nu$ by replacing $\chi(u)$ by 
$\chi_\text{SM}(u)$ on the right hand side of Eq.~(\ref{weinberg_eqn}). 
Here $\chi_\text{SM}(u)$ is the solution of Eq.~(\ref{weinberg_eqn}) for 
$f_\nu = f_\nu^\text{SM}$. Having made this approximation, 
Eq.~(\ref{weinberg_eqn}) reduces to
 \beq
{\mathcal{F}(u, \partial_u)} \, \chi(u) = f_\nu \int_0^u \mathcal{I}(u,U) \, \chi'_\text{SM}(U) \, \dd U  \,.
 \eeq
 Noting that the right hand side of this equation is proportional to
${\mathcal{F}(u, \partial_u)}\, \chi_\text{SM}(u)$, we have
  \beq
{\mathcal{F}(u, \partial_u)} \!\left(\chi(u) - \frac{f_\nu}{f_\nu^\text{SM}} \chi_\text{SM}(u) \right) = 0  \,.
 \eeq
 Recalling that ${\mathcal{F}(u, \partial_u)} \chi_0(u) = 0$, it follows that this 
equation admits a solution of the form,
 \beq
 \chi(u) - \frac{f_\nu}{f_\nu^\text{SM}} \chi_\text{SM}(u) \propto \chi_0(u) \; .
 \eeq 
 Using the initial conditions to fix the constant of proportionality, we 
obtain an analytic solution for $\chi(u)$ valid for general $f_\nu$ that 
is applicable in the short-wavelength limit,
 \beq
\chi(u)=\left(1+\fr{f_\nu}{f_\nu^\text{SM}}(A_{\rm SM}-1)\right)\!  \fr{\sin u}{u} \,.
 \eql{shortwave_chi}
 \eeq
 As discussed in~\cite{Weinberg:2003ur}, the ratio of the CMB 
tensor correlation functions for general $f_\nu$ \emph{relative to} that for 
$f_\nu^\text{SM}$ is given by 
 \beq
R = \left| \fr{\chi'(u)}{\chi'_\text{SM}(u)} \right|^2 
= \frac{\left( 1+\fr{f_\nu}{f_\nu^\text{SM}}(A_{\rm SM}-1) \right)^{\!\!2}}{A_{\rm SM}^2} \,.
 \eeq
 Since the right hand side of this equation is independent of $u$, it 
follows that the fractional change in the correlation functions arising 
from the presence of DR is independent of wave number for modes deep 
inside the horizon. The result is plotted in Fig.~2.

To obtain the solution of Eq.~(\ref{weinberg_eqn}) for \textit{long wavelengths}, 
it is convenient to change variables from $u$ to $y \equiv 
a(\tau)/a(\tau_\text{eq})$, where $a(\tau)$ is the standard scale factor 
of the Robertson-Walker metric and $\tau_\text{eq}$ is the value of the 
conformal time coordinate $\tau$ at matter-radiation equality. Expressed 
in terms of $y$, the integro-differential equation 
Eq.~(\ref{weinberg_eqn}) takes the form
 \beq
\left[ \mathcal{G}(y, \partial_y) + Q^2 \right] \chi(y) = f_\nu \int_0^y \mathcal{J}(y,Y) \, \chi'(Y) \, \dd Y
\,,
 \eql{weinberg_eqn} 
 \eeq
 Here $\mathcal{G}(y, \partial_y)$ is again a second-order, linear 
differential operator, and $Q$ is a normalized comoving wave number 
defined as $Q \equiv \sqrt2 \, k / k_\text{eq}$, where $k_\text{eq}$ is 
the value of $k$ corresponding to the length scale that enters the 
horizon at matter-radiation equality. We have also eliminated 
$\bar{f}_\nu(y)$ in favor of $f_\nu$, which is independent of $y$. The 
initial conditions are now given by $\chi(0) = 1$ and $\chi'(0) = 0$. 
The precise forms of $\mathcal{G}$ and the kernel function 
$\mathcal{J}(y,Y)$ may be found in~\cite{Weinberg:2003ur}. Then, given 
the solutions $\chi_0(y)$ and $\chi_\text{SM}(y)$ corresponding to the 
choices ${f}_\nu = 0$ and ${f}_\nu = {f}_\nu^\text{SM} (y)$, an 
approximate solution $\chi(y)$ corresponding to a general $f_\nu$ may be 
obtained as
 \beq
\chi(y) 
= 
\chi_0(y) 
+ \fr{f_\nu}{f_\nu^\text{SM}} \Bigl( \chi_\text{SM}(y) - \chi_0(y) \Bigr)
\,.
\label{general_chi_y}
 \eeq
 Expressions for $\chi'_\text{SM}(\ydec)$ and $\chi'_0(\ydec)$ at small 
$Q$ were obtained in~\cite{Stefanek:2012hj},
 \beq
\chi'_\text{SM}(\ydec)
&=
a_2 Q^2 + a_4 Q^4 + a_6 Q^6 + \cO(Q^8) 
\,,\\
\chi_0'(\ydec)
&=
b_2 Q^2 + b_4 Q^4 + b_6 Q^6 + \cO(Q^8).
 \eeq
 with
 \beqa{4}
a_2 &= -0.573661
\,,&\quad
a_4 &= 0.243294
\,,&\quad
a_6 &= -0.0381643
\,,\\
b_2 &= -0.601254
\,,&\quad
b_4 &= 0.264482
\,,&\quad
b_6 &= -0.0424186
\,.\nn
 \eeqa
  Here $\ydec$ is the value of $y$ at photon decoupling, given by $\ydec 
= 3.31$.
 Note that the regime of validity of these expansions is limited to 
small $Q$, and they are expected to break down at $Q\sim1$. Nonetheless 
these analytic approximations offer a simple way to parametrize the 
effects of DR on the tensor mode anisotropies for long wavelengths. 
Using these expressions in association with the general 
solution Eq.~(\ref{general_chi_y}), we can determine the suppression of the 
tensor correlation function for arbitrary $f_\nu$ relative to that for 
the SM,
 \beq
 R=
\left|\fr{\chi'(y_{\rm dec})}{\chi'_\text{SM}(y_{\rm dec})} \right|^2.
 \eeq
 The results are shown in Fig.~\ref{fig: BB_Q}, where we can see that 
scattering and free streaming DR contribute with opposite signs. It 
follows that while scattering DR tends to reduce the damping effect 
associated with the SM neutrinos, free streaming DR enhances this 
effect. This dissimilar behaviour is analogous to what was observed for 
the corrections to the amplitude and phase of the scalar modes.

\begin{figure}[t]
 \includegraphics[height=50mm]{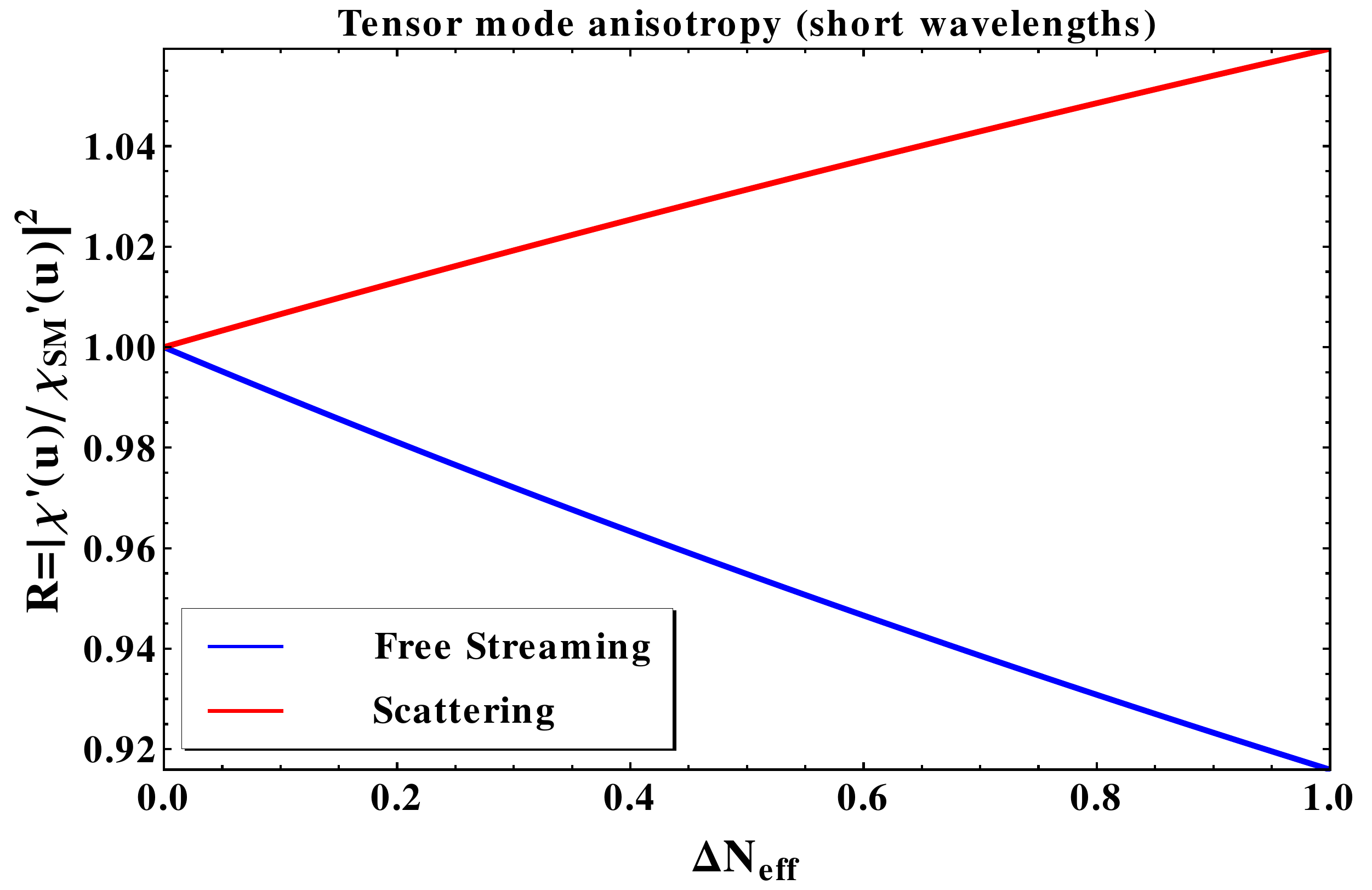} 
  \caption{ Relative change in tensor mode anisotropy compared to the SM case ($\Neff=3.046$), $R=\left|\fr{\chi'(u)}{\chi'_\text{SM}(u)} \right|^2$, for short wavelengths. }
\label{fig: tensor_short}
\end{figure}

\begin{figure}[t]
 \includegraphics[height=42mm]{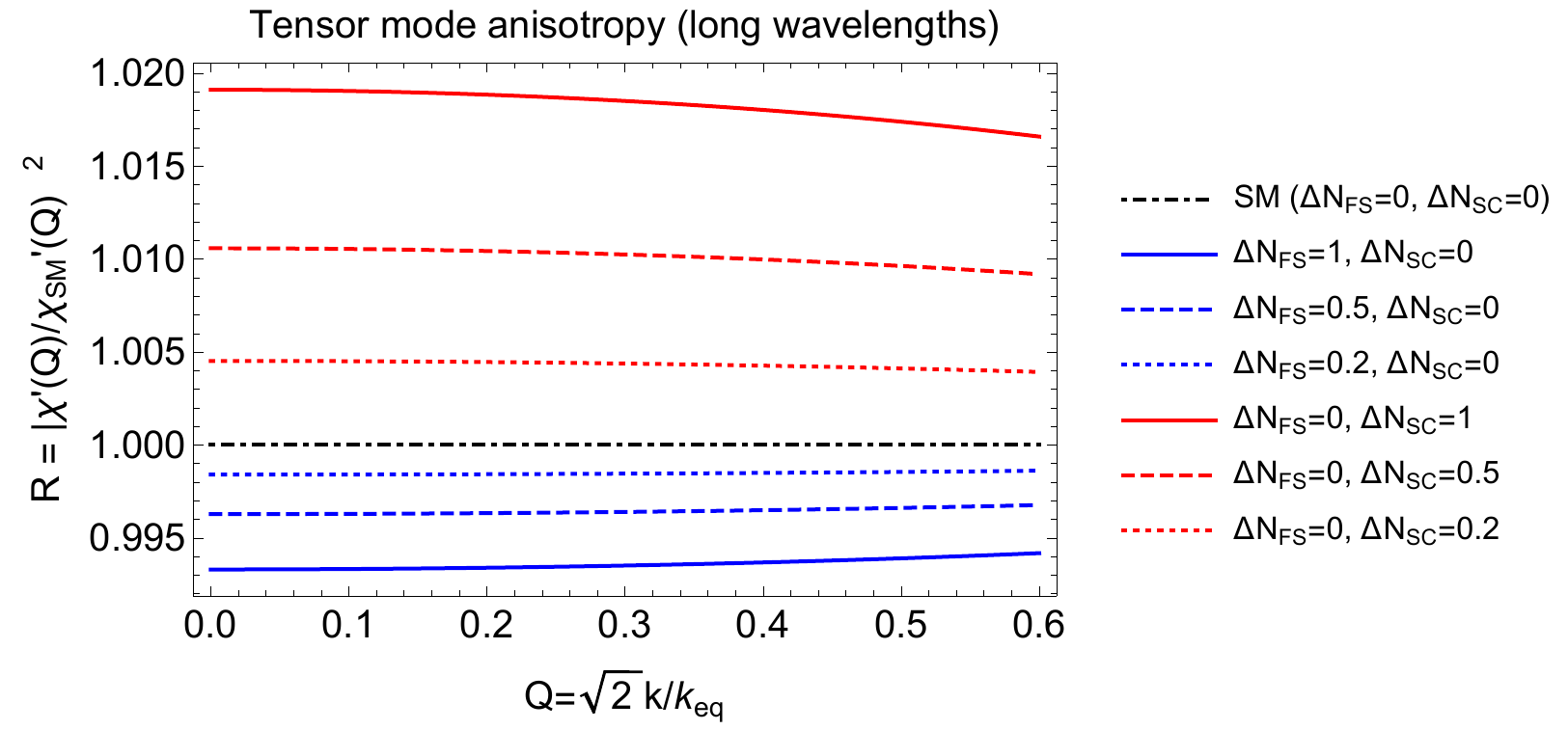} 
  \caption{ Relative change in tensor mode anisotropy compared to the SM case ($\Neff=3.046$), $R=\left|\fr{\chi'(y_{\rm dec})}{\chi'_\text{SM}(y_{\rm dec})} \right|^2$, for varying $\Delta N_{\rm eff}^{\rm free}$ (shorthand $\Delta N_{\rm FS}$) and $\Delta N_{\rm eff}^{\rm scatt}$ (shorthand $\Delta N_{\rm SC}$), for long wavelengths.}
\label{fig: BB_Q}
\end{figure}

\section{Determination of the Relic Abundance}
\label{sec:relic}


In this section we explain how the well-known formalism for determining 
the relic abundance of DM generalizes to the case when the DM sector and 
the SM are at two different temperatures. The evolution of the DM 
density in an expanding universe is governed by the Boltzmann equation,
 \beq
\frac{\mathrm{d}n_\chi}{\mathrm{d}t}+3Hn_\chi=-\langle \sigma v \rangle\left(n_\chi^2-(n_\chi^{\rm eq})^2  \right).
\label{eq:boltzmann}
 \eeq
 The difference between the scenario we are considering and the 
conventional relic abundance calculation for a thermal WIMP lies in the 
fact that in Eq.~(\ref{eq:boltzmann}), the averaged annihilation cross 
section $\langle \sigma \, v \rangle$ and the DM equilibrium number 
density $n_\chi^{\rm eq}$ now depend on the dark sector 
temperature $\hat{T}$ rather than the SM temperature $T$, while the 
Hubble constant $H$ depends on both $T$ and $\hat{T}$. In our analysis 
we will assume that at temperatures in the neighborhood of the 
freeze-out, the number of relativistic degrees of freedom in the SM, and 
in the dark sector, are not changing. Then, since the entropy densities 
of the two sectors are separately conserved, it follows that during 
freeze-out the ratio of the dark sector temperature $\hat{T}$ to the SM 
temperature $T$ is a constant labelled by $r \equiv \hat{T}/T$. We 
can then express all the terms in the Boltzmann equation as functions of 
the SM temperature $T$. For the equilibrium number density in the 
nonrelativistic limit we have,
 \begin{eqnarray}
n_\chi^{\rm eq} (\hat{T}) &=& g_{\chi} \!\left(\frac{m_{\chi} \hat{T}}{2 \pi}\right)^{\!\!\frac{3}{2}} \mathrm{e}^{- \frac{m_{\chi}}{\hat{T}}} 
\nonumber \\ 
&=&  g_{\chi} \!\left(\frac{r m_{\chi} {T}}{2 \pi}\right)^{\!\!\frac{3}{2}} \mathrm{e}^{- \frac{m_{\chi}}{r{T}}}.
 \eea
 For the Hubble constant,
\begin{align}
H^2 &= \dfrac{8 \pi G}{3} \rho =  \frac{8 \pi}{3 \Mpl^2} \left( \rho_{\text{SM}} + \rho_{\text{dark}} \right) \nonumber\\
&= \frac{8 \pi}{3 \Mpl^2} \left( \frac{\pi^2}{30} g_*(T) \, T^4 + \frac{\pi^2}{30} \hat{g}_*(\hat{T}) \, {\hat{T}}^4 \right) \nonumber\\
&\equiv  \frac{4 \pi^3}{45 \Mpl^2} g_{*\text{eff}} T^4
\,,
\end{align}
 where
 \begin{equation}
g_{*\text{eff}} (T, \hat{T}) \equiv g_* (T) + \hat{g}_*(\hat{T}) \!\left( \frac{\hat{T}}{T} \right)^{\!\!4} = g_*(T) + \hat{g}_*(\hat{T}) \, r^4.
 \end{equation}
 In the nonrelativistic regime $\langle \sigma v \rangle $ is well
approximated by
 \beq
\langle \sigma v \rangle\ = a + b \langle v^2 \rangle \,
\label{ab}
 \eeq 
 where the constant term plays the leading role if the annihilation can 
proceed through the s-wave. If, however, the s-wave contribution is 
suppressed so that annihilation occurs primarily through the p-wave, the 
term proportional to $v^2$ dominates. Performing the thermal average,
 \beq
a + b \langle v^2 \rangle = 
a + 6 r b \frac{{T}}{m_{\chi}}  \,.
\label{sigmav}
 \eeq  

 We are now in a position to solve the Boltzmann equation by the 
standard procedure. We introduce the variable $Y \equiv n_{\chi}/s_{\rm 
SM}$, where $s_{\rm SM}$ is the entropy density carried by the SM 
degrees of freedom. We also introduce the variable $x \equiv 
m_{\chi}/T$. In the radiation dominated era, when $g_*$ and $\hat{g}_*$ 
are not changing, we have
 \beq
\frac{\mathrm{d} x}{\mathrm{d}t} = H x
 \eeq  
 Eliminating  the number density $n_\chi$ in favor of $Y$, and $t$ in favor of $x$,
we obtain for the Boltzmann equation,
 \beq
\frac{\mathrm{d}Y}{\mathrm{d}x} = - \frac{\langle \sigma  v \rangle}{H x} s_{\rm SM} \left( Y^2 - Y_{\text{eq}}^2 \right)
 \eeq
 Rewriting this in terms of $\Delta \equiv Y - Y_{\text{eq}}$, we have
 \beq
\Delta' = - Y_{\text{eq}}' - f(x) \, \Delta \left( 2 Y_{\text{eq}} + \Delta \right)
 \eeq
 where $f(x)$ is given by
 \beq
f(x) 
\equiv \frac{\langle \sigma v \rangle s_\text{SM}}{Hx}
= 
\sqrt{\frac{\pi}{45}} \frac{g_*}{\sqrt{g_{*\text{eff}}}} m_{\chi} \Mpl \frac{a + 6 r b / x}{x^2}
 \eeq
 This Boltzmann equation cannot be solved exactly. However, it is 
possible to obtain analytic solutions of this equation that are valid at 
very early times, and very late times. Then, an approximate solution 
that is valid for all times may be obtained by stitching together these 
limiting cases. We define the freeze-out temperature $T_f$ in terms of the 
implicit relation,
 \beq
\Delta(x_f) = c Y_{\text{eq}}(x_f) \; ,
\label{eq:definexF}
 \eeq 
 where $c$ is an order one number. The final result depends only 
logarithmically on the value of $c$, which is chosen to be $\sqrt{2} - 
1$ in the case of s-wave annihilation and $\sqrt{3} - 1$ in the case of 
p-wave annihilation, to give the best fit to numerical results \cite{Kolb:1990vq}.  At early times $x \ll x_f$, $\Delta' \ll Y_{\text{eq}}'$, 
so that the Boltzmann equation reduces to
 \beq
\Delta = - \frac{Y_{\text{eq}}'}{f(x) \left( 2 Y_{\text{eq}} + \Delta \right)}
 \eeq
 This equation, in combination with Eq.~(\ref{eq:definexF}), may be used to 
determine the freeze-out temperature,
 \beq
 x_f = r \log \left( \fr{c(c+2)}{4 \pi^3} \sqrt{\frac{45}{2}} \frac{g_{\chi}}{\sqrt{g_{*\text{eff}}}}  m_\chi \Mpl \frac{r^\frac{5}{2} \!\left( a + 6 r b / x_f \right)}{x_f^{1/2}}  \right)
 \label{x_f}
 \eeq
 At late times, $x \gg x_f$, $\Delta \approx Y \gg Y_{\text{eq}}$ and $\Delta' \gg 
Y_{\text{eq}}'$ so that we have,
 \beq
Y^{-2} Y' = - f(x)
 \eeq
 Integrating this equation from $x_f$ to $\infty$, we obtain
 \beq
Y_{\infty}= \left[\sqrt{\frac{\pi}{45}} \frac{g_*} {\sqrt{g_{*\text{eff}}}}  \Mpl \; m_{\chi} \, \fr{a + 3 r b / x_f}{x_f} \right]^{-1} \label{eq:DMdensity}
 \eeq  
 where we have used the fact that $\Delta(x_f) \gg \Delta(\infty)$. 
Combining this result with the expression for $x_f$ from 
Eq.~(\ref{x_f}), we can obtain the present day energy density in dark 
matter, $\Omega_{\chi} = m_{\chi} s_{0} Y_{\infty}\rho_c^{-1}$. 
Here $s_0$, $\rho_c$ are the present-day SM entropy density and critical density, respectively. 

From this discussion we see that the difference between our framework, 
which allows for $\hat{T} \neq T$ at freeze-out, and the conventional 
scenario with $\hat{T} = T$, primarily translates into an extra factor 
of $r = \hat{T}/T$ in the expression for the relic abundance of DM\@. 
There is an additional effect arising from the $r$ dependence of the 
argument of the logarithm in Eq.~({\ref{x_f}}), but this is small. Put 
in another way, the values of $a$ and $b$ in the expression for $\langle 
\sigma v \rangle$ in Eq.~(\ref{ab}) that correspond to the observed 
$\Omega_{\rm DM}$ are smaller by a factor of $\hat{T}/T$ than in the 
case of the standard thermal cross section $\sim 3 \times 10^{-26}\,\rm 
cm^3/s$. One may obtain similar results to those given in 
Eqs.~(\ref{x_f}) and~(\ref{eq:DMdensity}) by performing a simple 
estimate based on equating $\Gamma_\chi=n_\chi^{\rm eq} \langle \sigma v 
\rangle$ to the Hubble constant $H$ at freeze-out, while keeping track 
of the distinction between $T$ and $\hat{T}$.

In this analysis, we have assumed that the number of degrees of freedom 
in the SM, and in the hidden sector, do not change in the neighborhood 
of the freeze-out temperature. We have further assumed that the 
annihilation cross section is well approximated by an expansion of the 
form given in Eq.~(\ref{sigmav}). Using the methods 
of~\cite{Srednicki:1988ce, Gondolo:1990dk}, these assumptions can be 
relaxed, and the result generalized. We leave this for future work.


\section{A Simple Benchmark Model}


\subsection{The Model}

In this section we consider in detail a simple model that illustrates 
the scenario we are considering. We consider an unbroken U(1) gauge 
theory with massless spin-$1/2$ fermions charged under it. Let 
$\hat{A}_\mu$ be the massless gauge boson associated with this U(1), and 
$\hat{\psi}_i$ and $\hat{e}_i$ ($i = 1, 2, \cdots, N_{\hat{\psi}}$) 
represent the massless (4 component) fermions and their associated 
charges under U(1)$_{\hat{A}}$. Note that some of the $\hat{e}_i$ may be 
zero, thereby allowing us to dial the number of free-streaming species.  
We consider a complex scalar particle, denoted by $\chi$, as the DM 
candidate.
  
Since our scenario assumes hidden sector dark matter, $\chi$ is 
uncharged under the SM gauge interactions. The only possible 
renormalizable interaction of $\chi$ with SM particles is of the Higgs 
portal form $|\chi|^2 |H|^2$, where $H$ is the complex scalar doublet 
that includes the SM Higgs particle $h$. This operator is expected to be 
present, and serves to ensure that the SM and hidden sectors are 
initially in thermal and chemical equilibrium. However, our interest is 
in the scenario in which this coupling is not large enough to govern the 
annihilation of dark matter that determines its relic abundance. 
Instead, the relic abundance is controlled by an additional interaction 
through which dark matter annihilates into DR\@. For that purpose, we 
introduce a massive spin-1 boson $\hat{Z}$ that couples to both $\chi$ 
and $\hat{\psi}_i$, which carry charges $\hat{q}_\chi$ and $\hat{q}_i$ 
under the associated broken U(1)$_{\hat{Z}}$. In our analysis, we will 
focus on the parameter range where $\chi$ is lighter than the $\hat{Z}$ 
gauge boson by a factor of a few.

The Lagrangian of our benchmark model therefore reads as
 \begin{equation}
\begin{alignedat}{3}
\mathcal{L} 
=\>& 
\mathcal{L}_\text{SM}
+(\mathrm{D}_\mu\chi)^* (\mathrm{D}^\mu \chi) 
-\hat{m}_\chi^2 |\chi|^2
-\kappa |\chi|^2 |H|^2  
-\frac{\lambda}{4} |\chi|^4 
\\
&
+\bar{\hat{\psi}}^i \mathrm{i} \slashed{\mathrm{D}} \hat{\psi}_i
-\frac14 \hat{Z}_{\mu\nu} \hat{Z}^{\mu\nu} + \frac12 m_{\hat{Z}}^2 \hat{Z}^2 
-\frac14 \hat{A}_{\mu\nu} \hat{A}^{\mu\nu}
\\
&+\cdots
\end{alignedat}
\label{eq:the_model}
 \end{equation}
 In this expression the ellipses represents the sector responsible for 
breaking U(1)$_{\hat{Z}}$ and generating the $\hat{Z}$ mass. We will not 
specify this sector explicitly as it is not relevant for our 
discussions. We only mention that all the particles in that sector are 
assumed to be significantly heavier than $m_\chi$ and $m_{\hat{Z}}$, so 
that they do not affect the dynamics we are considering.%
\footnote{ 
 We assume that the Lagrangian in Eq.~(\ref{eq:the_model}) does not 
include kinetic mixings between these new U(1) gauge fields and 
those of the SM, such as $\hat{A}_{\mu\nu} B^{\mu\nu}$, where $B_\mu$ is 
the SM hypercharge gauge boson. These can be forbidden by an internal 
charge conjugation symmetry carried by the hidden sector fields.}
Note that, as a consequence of electroweak symmetry breaking, the 
parameter $\hat{m}_\chi$ in the Lagrangian is not equal to the $\chi$ 
mass, $m_\chi$, but is related to it as
 \begin{equation}  
m_\chi^2 = \hat{m}_\chi^2 + \frac{\kappa}{2} v_\text{\tiny EW}^2
\,,
\end{equation}
 where $v_\text{\tiny EW} = 246$~GeV\@. The masses of the $\hat{\psi}_i$ are all 
set to zero, and are protected against quantum corrections by chiral 
symmetry. Finally, to ensure the stability of the dark matter particle 
$\chi$, we impose an exact $\mathbbm{Z}_2$ symmetry under which $\chi$ 
is odd and all the other fields are even.%
\footnote{
Alternatively, such $\mathbbm{Z}_2$ symmetry can accidentally emerge if 
the value of $\hat{q}_\chi$ is such that it is not equal to any linear 
combination of the other U(1)$_{\hat{Z}}$ charges with rational coefficients.}

In this benchmark model, there is only one renormalizable interaction, $ 
\kappa |\chi|^2 |H|^2$, that connects the dark sector to the SM\@. The 
presence of such an interaction is expected to be a general feature of 
any theory in which the dark matter particle is a scalar, as there is no 
symmetry that forbids it. Provided $\kappa \gtrsim 10^{-6}$, this 
coupling ensures that the SM and hidden sector are initially in thermal 
equilibrium at temperatures of order the weak scale. The requirement 
that annihilation to SM states does not play a significant role in 
setting the relic abundance constrains $\kappa \lesssim 10^{-2}$. We 
will therefore focus on the regime $10^{-2} \gtrsim \kappa \gtrsim 
10^{-6}$. We also choose to stay away from the region of parameter space 
such that the DM mass is close to half the Higgs mass, where 
annihilation to the SM is resonantly enhanced.

The presence of interactions between the hidden sector and the SM leads 
to constraints on the theory and potential signals. Upon electroweak symmetry breaking, the 
$|\chi|^2 |H|^2$ term leads to a 3-point interaction of the form $h 
|\chi|^2$. Provided $m_\chi < m_h / 2$, this will generate 
a contribution to the invisible width of the Higgs boson given by
 \begin{equation} 
\Gamma_h^\text{inv} = \frac{m_h}{16\pi} \frac{\kappa^2 v_\text{\tiny EW}^2}{m_h^2} \, \beta_\chi \,,
\quad \beta_\chi \equiv \sqrt{1 - \dfrac{4m_\chi^2}{m_h^2}} \,. 
 \end{equation} 
  The current experimental limit on the invisible branching ratio from 
the LHC 7 and 8 TeV datasets stands at about $30 \%$ of the total Higgs 
width, $\Gamma_{h} = 4.07 \times 10^{-3}$ GeV, from the vector boson 
fusion channel~\cite{ATLASVBF, CMSVBF}. The limit from the associated 
production channel is significantly weaker~\cite{ATLASAP, CMSAP}. The 
bound on the invisible width is expected to improve to $\sim 10 \%$ of 
the total width after 300 fb$^{-1}$ at the 14 TeV 
LHC~\cite{ATLAS:2013hta, CMS:2013xfa}. For $m_h \gg m_{\chi}$, these 
limits can be translated into the constraints on the parameter $\kappa$.  
At present we have $\kappa \lesssim 2 \times 10^{-2}$ from the LHC 7 and 
8 TeV runs. This limit is expected to improve 
to $\kappa \lesssim 7 \times 10^{-3}$ after 300 fb$^{-1}$ at the 14 TeV 
LHC\@. Stronger bounds can be obtained at future lepton colliders, such as 
the ILC or TLEP. According to the studies~\cite{Dawson:2013bba, 
Chacko:2013lna,Han:2013kya} these machines can constrain the Higgs 
invisible branching ratio to $0.2 \% \sim 1 \%$, which corresponds to 
$\kappa \lesssim 9.5 \times 10^{-4} \sim 2.5 \times 10^{-3}$. For $m_h < 
2 m_{\chi}$, on-shell Higgs decays into DM particles are kinematically 
forbidden, and the hidden sector must be accessed through an off-shell 
Higgs, or through loop effects. The current collider limits are then 
very weak in the regime of interest, $\kappa \lesssim 10^{-2}$, and are 
not expected to improve significantly even at future 
colliders~\cite{Chacko:2013lna}. While the $h|\chi|^2$ interaction can 
also give rise to signals in direct/indirect DM detection experiments, 
current experiments are not yet sensitive in the region 
$\kappa \lesssim 10^{-2}$.

 \subsection{Relic Abundance of DM}

Let us now analyze this benchmark model. The first step is to require 
that the relic abundance of $\chi$ agree with the observed amount of 
cold dark matter. We focus on the region of parameter space where the DM 
mass lies between 5 GeV and 100 GeV, and where $2m_\chi < m_{\hat{Z}}$. 
Then the abundance of DM is governed by the annihilation process $\chi + 
\chi^* \to \hat{\psi} + \bar{\hat{\psi}}$ via the exchange of a virtual 
$\hat{Z}$ in the $s$-channel, after the DM particles become 
nonrelativistic. To lowest order in the nonrelativistic limit, the cross 
section for this process is given by
 \begin{equation}
\sigma_{\chi\chi^* \to \hat{\psi}\bar{\hat{\psi}}} 
= 
\frac{\hat{g}^4_\text{eff}}{48 \pi m_\chi^2}
\!\left( 1 - \frac{m_{\hat{Z}}^2}{4 m_\chi^2} \right)^{\!\!-2} \!
v_\chi
\,,
 \end{equation}
 where
 \begin{equation}
\hat{g}^4_\text{eff} \equiv \sum_{i=1}^{N_{\hat{\psi}}} \hat{g}_{\hat{Z}}^4 \hat{q}^2_\chi \hat{q}^2_i
 \end{equation}
 and $v_\chi$ represents the speed of each annihilating $\chi$ in the 
center-of-momentum (CM) frame. The presence of the $v_\chi$ suppression is an 
indication that the annihilation process proceeds through the $p$-wave 
channel. This can be understood from the fact that the initial state 
consists of two scalars and the intermediate state is spin-1. Letting 
$v$ be the relative velocity of the annihilating $\chi$'s ($v 
=2v_\chi$), and performing a thermal averaging, we obtain
 \begin{equation}
\langle \sigma_{\chi\chi^* \to \hat{\psi}\bar{\hat{\psi}}} \, v \rangle
= 
\frac{\hat{g}^4_\text{eff}}{48 \pi m_\chi^2}
\!\left( 1 - \frac{m_{\hat{Z}}^2}{4 m_\chi^2} \right)^{\!\!-2}
\frac{3 \hat{T}}{m_\chi}
\,.\eql{sigma_v:chi_annihilation}
 \end{equation}
 In this expression $\hat{T}$ is the temperature of the dark sector and 
we have used the nonrelativistic relation $\langle v^2 
\rangle = 6 \hat{T} / m_{\chi}$. We emphasize again that $\hat{T}$ is in general no 
longer equal to the temperature $T$ of the SM gas, because we are 
allowing for the possibility that the dark sector decouples from the SM 
well before before the $\chi$ particles freeze out.

 The relic abundance of $\chi$ can be obtained from the Boltzmann equation,
 \beq
\frac{\mathrm{d}n_\chi}{\mathrm{d}t}+3Hn_\chi=-\langle \sigma_{\chi\chi^* \to \hat{\psi}\bar{\hat{\psi}}} 
\, v \rangle\left(n_\chi^2-(n_\chi^{\rm eq})^2  \right).
\label{eq:boltzmann2}
 \eeq
 This can be solved using the methods discussed in section~\ref{sec:relic}.

 \subsection{Kinetic Decoupling between DM and DR}

After freeze-out, the system of DM and DR continues to behave as a 
tightly coupled fluid until kinetic decoupling occurs. This is in 
analogy to the photon-baryon plasma before recombination, and the DM-SM 
plasma in the context of the conventional WIMP scenario. Even though DM 
and DR have chemically decoupled, they are kept in kinetic equilibrium 
by the elastic scattering of $\chi$ with the fermions $\hat{\psi}_i$, which 
proceeds through an off-shell $\hat{Z}$ in the t-channel. To leading 
order in the nonrelativistic limit, the cross section for $\chi + \hat{\psi}_i 
\to \chi + \hat{\psi}_i$ is given by
 \begin{equation}
\sigma_{\chi\hat{\psi}_i \to \chi\hat{\psi}_i} 
= \frac{\hat{g}_{Z}^4 \hat{q}^2_\chi \hat{q}^2_i}{2 \pi} \frac{E_{\hat{\psi}_i}^2}{m_{\hat{Z}}^4}
\,,
 \end{equation}
 where $E_{\hat{\psi}_i}$ represents the energy of the scattering $\hat{\psi}_i$ 
 in the CM frame.
Performing a thermal averaging, we obtain
 \begin{equation}
\sum_{i=1}^{N_{\hat{\psi}}}
n_{\hat{\psi}_i} \langle \sigma_{\chi\hat{\psi}_i \to \chi\hat{\psi}_i} v \rangle
= 
\frac{4 \cdot 45\zeta(5)}{4 \pi^2} \frac{\hat{g}^4_\text{eff}}{2 \pi} \frac{\hat{T}^5}{m_{\hat{Z}}^4}
\equiv \Gamma_{\rm col}
\,,
 \end{equation}
 where the overall factor of $4$ in front accounts for the two 
polarizations of $\hat{\psi}$, as well as the contribution from scattering 
with an anti-$\hat{\psi}_i$. The factor of $45 \zeta(5) / 4\pi^2$ arises from 
thermal averaging. Note that $\Gamma_{\rm col}$ represents the rate for a $\chi$ 
to experience a single collision with any one of the $\hat{\psi}_i$. This rate 
is not, however, the same as the rate for the $\chi$ plasma to thermalize 
with the $\hat{\psi}$ bath. This is because a single collision of a $\chi$ 
particle with a $\hat{\psi}_i$ typically involves a momentum transfer of only 
$\mathcal{O}(\hat{T})$, which is not large enough to significantly 
deflect the direction of the $\chi$, which carries a momentum 
$\mathcal{O}(\sqrt{m_\chi \hat{T}})$. Viewing the effect of $N$ such 
collisions as a random walk with $N$ steps, the $\chi$ momentum 
typically changes by $\mathcal{O}(\sqrt{N} \hat{T})$ after $N$ 
collisions. Requiring that this change be $\mathcal{O}(\sqrt{m_\chi 
\hat{T}})$, we obtain that $N \sim m_\chi / \hat{T}$. It follows that 
the rate for the $\chi$ plasma to thermalize with the $\hat{\psi}$ bath is given 
by
 \begin{equation}
\frac{\Gamma_{\rm col}}{N} 
\sim 
\frac{45\zeta(5)}{\pi^2} \frac{\hat{g}^4_\text{eff}}{2 \pi} \frac{\hat{T}^6}{m_{\hat{Z}}^4 m_\chi}
\,.
 \end{equation}
 Then, the temperature $\hat{T}_\text{D}$ below which the $\chi$ plasma 
thermally decouples from the bath may be estimated as
 \begin{equation}
\frac{\Gamma_{\rm col}}{N}
\biggr|_{\hat{T} = \hat{T}_\text{D}}
\sim H\biggr|_{T = T_\text{D}}
\,.
 \end{equation}
 For the ranges of parameters considered in our benchmark models, a 
quick estimate tells us that $T_\text{D} \sim 
\mathcal{O}(1)$--$\mathcal{O}(10)$ MeV\@.

Above the kinetic decoupling temperature $T_\text{D}$, DM is in 
equilibrium with DR, and they form a tightly coupled fluid. Acoustic 
oscillations within this fluid have the effect of erasing density 
perturbations on small scales. As a result, the temperature $T_\text{D}$ 
determines the cutoff of the power spectrum, and sets a lower bound on 
the masses of the smallest halos~\cite{Loeb:2005pm}, 
 \beq
M_{\rm cut} \simeq& 10^{5} \!\left(\frac{T_{\text D}}{10~\rm keV} \right)^{\!\!-3}M_\odot.
 \eeq 
 In principle, this offers a separate way to probe these theories, 
independent of the CMB\@. At present, kinetic decoupling temperatures up 
to about 10 keV can be probed.  Unfortunately, in our scenario, 
$T_\text{D}$ is too high for the cutoff of short-distance structures to 
be observable in current experiments.

One possible generalization of our scenario that would lead to 
observable effects in the DM power spectrum involves allowing the mass 
$m_{\hat{Z}}$ of the mediator to lie below the weak scale. It is easy to 
see that the WIMP miracle prediction for DM relic abundance continues to 
apply, provided the DM candidate itself continues to have weak scale 
mass and couplings of $\mathcal{O}(1)$ to the mediator. In order to have 
observable effects, a mediator of mass $\hat{m}_Z \lesssim 
\mathcal{O}(100)$ MeV would be required. If the mediator mass is further 
pushed down to $O(\rm MeV)$, exchange of the light $\hat{Z}$ would give 
rise to sizable DM self-interactions, which can lead to observable 
effects in the DM power spectrum~\cite{Spergel:1999mh}. The resulting DM 
scattering cross section is velocity dependent, and can therefore be 
large enough to resolve small scale structure anomalies while remaining 
consistent with cosmological bounds, along the lines suggested 
in~\cite{Feng:2009mn, Loeb:2010gj, Aarssen:2012fx, Tulin:2012wi, 
Tulin:2013teo}. We leave a detailed study of this possibility for future 
work.

\subsection{Kinetic Decoupling of DM from the SM bath}

As we noted earlier, the size of the CMB signals depends on the 
temperature $T = \hat{T} = \Td$ at which the dark matter sector 
kinetically decouples from the SM bath. For $m_\chi $ in the range 
between 10 GeV and 100 GeV, the leading processes maintaining the 
kinetic equilibrium are the elastic scattering of $\chi$ off the $W$, 
$Z$, and SM fermions. These processes decouple at different 
temperatures, and $\Td$ is set by the last process that decouples.

It turns out that at temperatures above $\sim 10$~GeV 
the dominant process is the scattering of 
DM particles off $W$ bosons. 
At these temperatures, even if the $W$ bosons are nonrelativistic 
and their number density begins to be Boltzmann suppressed, 
it is still more dominant than the competing scattering of $\chi$ off a relativistic $b$ quark,
as the latter suffers from a small Yukawa coupling. 
Let's first consider the scattering of a relativistic $\chi$ off a nonrelativistic $W$. 
The tree-level cross section in this limit is given by
\beq
\sigma_\W = \fr{\kappa^2}{4\pi} \fr{m_\W^2}{m_h^4} 
\,.
\eeq
Equating the scattering rate $\Gamma_W=n_\W \langle \sigma_\W v 
\rangle$ with the Hubble expansion rate $H$
gives an estimate for the 
temperature at which this process decouples, where $n_\W$ represents the 
equilibrium number density of $W$ bosons, and the relative speed $v$ between the scattering $\chi$ and $W$ is unity in the limit under consideration. We thus obtain
\beq
\Td 
&\sim 
m_\W \!\left[ \log\!\left( 
\fr{9\sqrt{10}}{16\pi^4} \fr{\MP}{m_h} \fr{m_\W^3}{m_h^3} \fr{\kappa^2}{\sqrt{g_*}}
\right) \right]^{-1}
\\
&= 
m_\W \!\left[ 8.5 + \log\!\left( \fr{\kappa^2}{10^{-10}} \fr{10}{\sqrt{g_*}} \right) \right]^{-1}
\,.\eql{T_kd:W:R-NR}
\eeq
 In obtaining this expression, we have assumed that $g_* \gg \hat{g}_*$, 
so that the energy density is dominated by the SM degrees of freedom. 
For a similar process with the $Z$ boson rather than the $W$, every 
appearance of $m_\W$ above should be replaced by $m_\Z$, and the argument of
the logarithm is reduced by a factor of 2 because of the fewer
degrees of freedom associated with the $Z$. Therefore, the corresponding 
decoupling temperature is proportional to $m_\Z$ with the same 
proportionality factor (neglecting the change in the argument of the logarithm), so the $Z$ decouples at a slightly higher 
temperature than the $W$.

On the other hand, in the limit where the $\chi$ and $W$ are both nonrelativistic, 
the scattering cross section is given at tree level by
\beq
\sigma_\W 
=
\fr{\kappa^2}{4\pi (m_\chi + m_\W)^2} \fr{m_\W^4}{m_h^4}
\,.\label{eq:sigma_W:NR-NR}
\eeq
To calculate the rate  
$n_\W \langle \sigma_\W v \rangle$, 
we need to know $\langle v \rangle$.
For a single nonrelativistic particle of mass $m$ obeying the Maxwell-Boltzmann distribution with temperature $T$, 
an elementary integral gives $\langle v \rangle = \sqrt{8T / (\pi m)}$.
We have a two-body system instead, but the relative speed $v$ in the lab frame is the same as that in the CM frame, 
where the two-body problem reduces to a one-body problem with the reduced mass $\mu_{\chi\W} \equiv m_\chi m_\W / (m_\chi + m_\W)$.
We thus have $\langle v \rangle = \sqrt{8T / (\pi \mu_{\chi\W})}$. 
The kinetic decoupling temperature then becomes
\beq
\Td 
&\sim 
m_\W \!\left[ \log\!\left( 
\fr{9\sqrt{5}}{4\pi^{9/2}} 
\fr{\MP}{m_h} \fr{m_\W^3}{m_h^3} 
\fr{\mu_{\chi\W}^2}{m_\chi^2} \sqrt{\fr{m_\W}{\mu_{\chi\W}}}
\fr{\kappa^2}{\sqrt{g_*}}
\right) \right]^{-1}
\\
&=
m_\W \left[ 
8.9
+\log\!\left(
\fr{\mu_{\chi\W}^2}{m_\chi^2} \sqrt{\fr{m_\W}{\mu_{\chi\W}}}
\fr{\kappa^2}{10^{-10}} \fr{10}{\sqrt{g_*}}
\right) \right]^{-1}
\,.\eql{T_kd:W:NR-NR}
\eeq

At temperatures below $\sim 10$~GeV, scattering of DM off SM fermions 
becomes increasingly important, and eventually dominates. 
At these temperatures the DM particle $\chi$ is nonrelativistic, 
while the SM fermions may be relativistic or nonrelativistic. 
In the case where the scattering SM fermion $f$ is relativistic, 
the tree-level cross section is given by
\beq
\sigma_f 
= \fr{\kappa^2}{8\pi m_\chi^2} \fr{m_f^2 p^2}{m_h^4} 
\,,
\eeq
where $p$ is the magnitude of the 3-momenta of the scattering particles in the CM frame, and higher order terms in $p / m_\chi$ have been neglected.
In this leading nonrelativistic approximation, 
the thermal average $\langle p^2 \rangle$ in the CM frame is the same as that in the lab frame, as the difference between the two frames itself is an $\cO(p/m_\chi)$ effect.
Therefore, we evaluate $\langle p^2 \rangle$ and the number density $n_f$ simply in the lab frame in the standard way, obtaining
\beq
\langle p^2 \rangle = \fr{15 \zeta(5)}{\zeta(3)} \, T^2
\,,\quad
n_f = 4 N_\mathrm{c} \!\cdot\! \fr{3 \zeta(3)}{4 \pi^2} \, T^3
\,,
\eeq
where $N_\mathrm{c}$ is the number of colors carried by the fermion species $f$.
Notice, however, that the relevant reaction rate to maintain a kinetic equilibrium is not simply given by $n_f \langle \sigma_f v \rangle$,
because a single collision of a heavy particle $\chi$ with a relativistic particle $f$ with $p \ll m_\chi$ hardly changes the direction of the $\chi$ momentum. 
Since the $\chi$ momentum is on average of order $\sqrt{m_\chi T}$,
and a typical momentum transfer via a single collision is of order $p \sim T$, 
the number of collisions $N$ required to alter the $\chi$ momentum by an $\cO(1)$ fraction is given by $\sqrt{N} T \sim \sqrt{m_\chi T}$, i.e., $N \sim m_\chi / T$. 
The relevant reaction rate to maintain a kinetic equilibrium, therefore, 
is given by $\Gamma_f = n_f \langle \sigma_f v \rangle / N$. Then, from $\Gamma_f \sim H$, we obtain an estimate for the kinetic decoupling temperature:
\beq
\Td 
&\sim 
m_h \!\left( 
\fr{16\pi^{9/2}}{135 \sqrt5 \,\zeta(5)} \fr{1}{N_\mathrm{c}}
\fr{m_\chi^3}{m_f^2 \MP} \fr{\sqrt{g_*}}{\kappa^2}
\right)^{\!\!\frac14}
\\
&\sim 
4.3\GeV \times \!\left( 
\fr{3}{N_\mathrm{c}}
\fr{m_\chi^3}{(10\GeV)^3} \fr{m_b^2}{m_f^2} 
\fr{10^{-10}}{\kappa^2} \fr{\sqrt{g_*}}{10}
\right)^{\!\!\frac14} 
,\eql{T_kd:f:R-NR}
\eeq
where $m_b$ is the $b$-quark mass.

\begin{figure}[t]
\centering
\includegraphics[height=58mm]{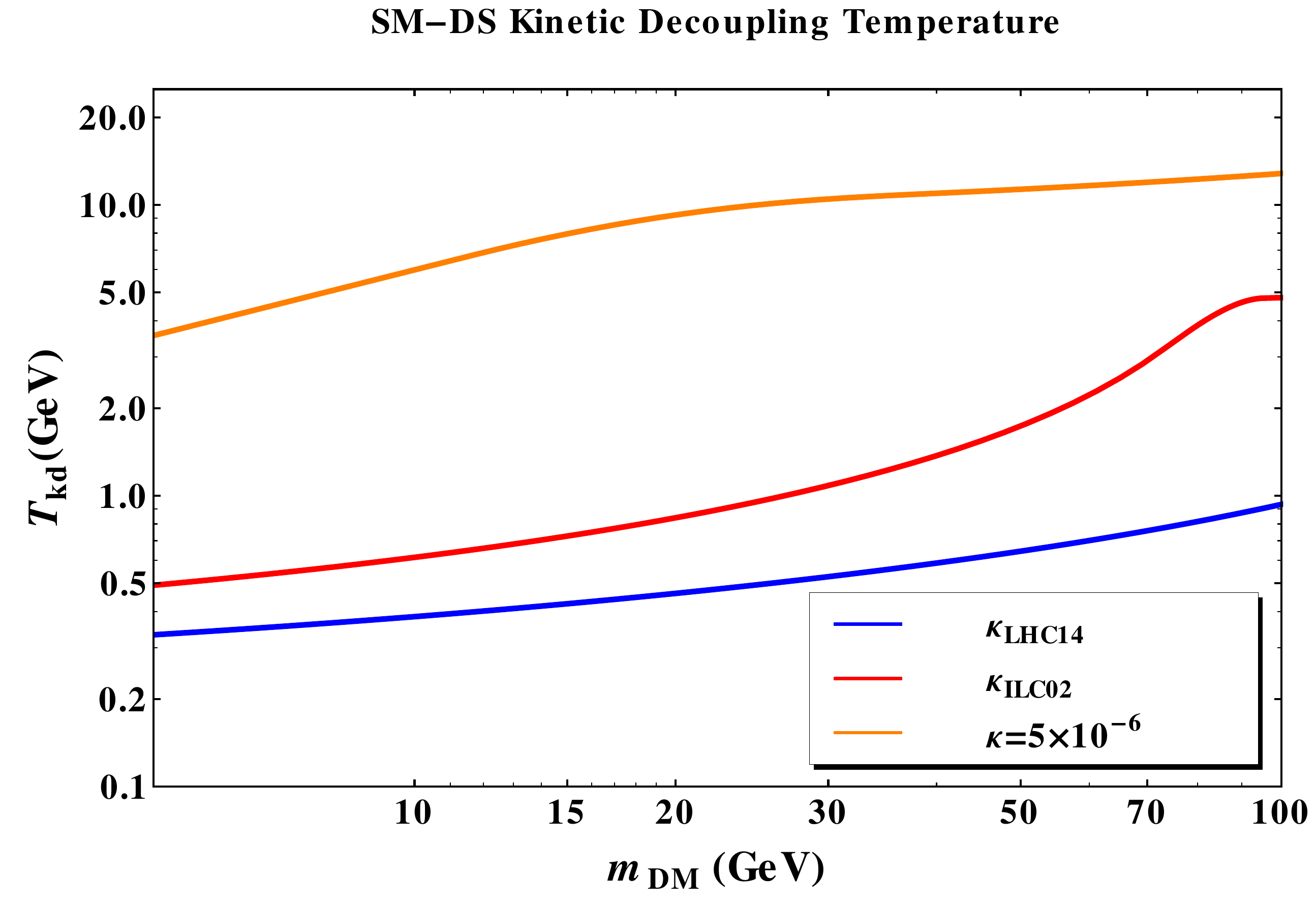}
\caption{The kinetic decoupling temperature $\Td$ between the SM and hidden sectors 
as a function of the DM mass $m_\text{DM}$ 
for three values of $\kappa$: $\kappa_\text{LHC14} = 7 \times 10^{-3}, \kappa_\text{ILC02} = 9.5 \times 10^{-4}$ and $\kappa = 5 \times 10^{-6}$.}
\label{fig:kinetic_decoupling_temperature_vs_m_DM}
\end{figure}

If the SM fermion is also nonrelativistic, 
the tree-level cross section in the leading nonrelativistic approximation 
is given by
\beq
\sigma_f 
=
\fr{\kappa^2}{4\pi (m_\chi + m_f)^2} \fr{m_f^4}{m_h^4}
\,.
\eeq
For thermal averaging, we use $\langle v \rangle = \sqrt{8T / (\pi \mu_{\chi f})}$, 
similarly to what we did below Eq.~(\ref{eq:sigma_W:NR-NR}).  
Furthermore, the number of collisions $N$ required to randomize the motion of $\chi$ 
is given by $\sqrt{N} \sqrt{m_f T} \sim \sqrt{m_\chi T}$, i.e., $N \sim m_\chi / m_f$.
Then, from $n_f \langle \sigma_f v \rangle / N \sim H$, we obtain
\beq
\Td 
&\sim 
m_f \!\left[ \log\!\left( 
\fr{3\sqrt{5}}{2\pi^{9/2}} 
\!\cdot\! N_\mathrm{c} \!\cdot\!
\fr{\MP}{m_h} 
\fr{m_f^6}{m_h^3 m_\chi^3} 
\!\left( \fr{\mu_{\chi f}}{m_f} \right)^{\!\!\frac32}\!\!
\fr{\kappa^2}{\sqrt{g_*}}
\right) \right]^{-1}
\\
&=
m_f \!\left[ 
5.2
+\log\!\left(\!\!
N_\mathrm{c} 
\!\left(\! \fr{100 \, m_f^2}{m_h m_\chi} \right)^{\!\!3} \!
\!\left(\! \fr{\mu_{\chi f}}{m_f} \right)^{\!\!\frac32}\!\!
\fr{\kappa^2}{10^{-6}} \fr{10}{\sqrt{g_*}}
\right) \!\!\right]^{\!-1} \!\!\!
.\eql{T_kd:f:NR-NR}
\eeq
Note that the ``reference value'' of $\kappa$ used here is $10^{-3}$, 
which differs from Eqs.~\eq{T_kd:W:R-NR}, \eq{T_kd:W:NR-NR}, and 
\eq{T_kd:f:R-NR}, where the value used was $10^{-5}$. This reflects the 
fact that this process is relevant only if $\kappa$ is sufficiently 
large, even for the $b$ quark.

In Fig.~\ref{fig:kinetic_decoupling_temperature_vs_m_DM}, 
we plot the 
kinetic decoupling temperature against the dark matter mass, for several 
different values of the coupling $\kappa$. We see that for the range of 
values of $\kappa$ that can be probed in current and future collider 
experiments, the kinetic decoupling temperature lies below several GeV and above several hundred MeV\@.
In particular, the kinetic decoupling occurs before a drastic change in the number of relativistic SM degrees of freedom due to QCD phase transition.

\subsection{Signals}

\begin{figure}[htb!]
\centering
\mbox{\includegraphics[height=50mm]{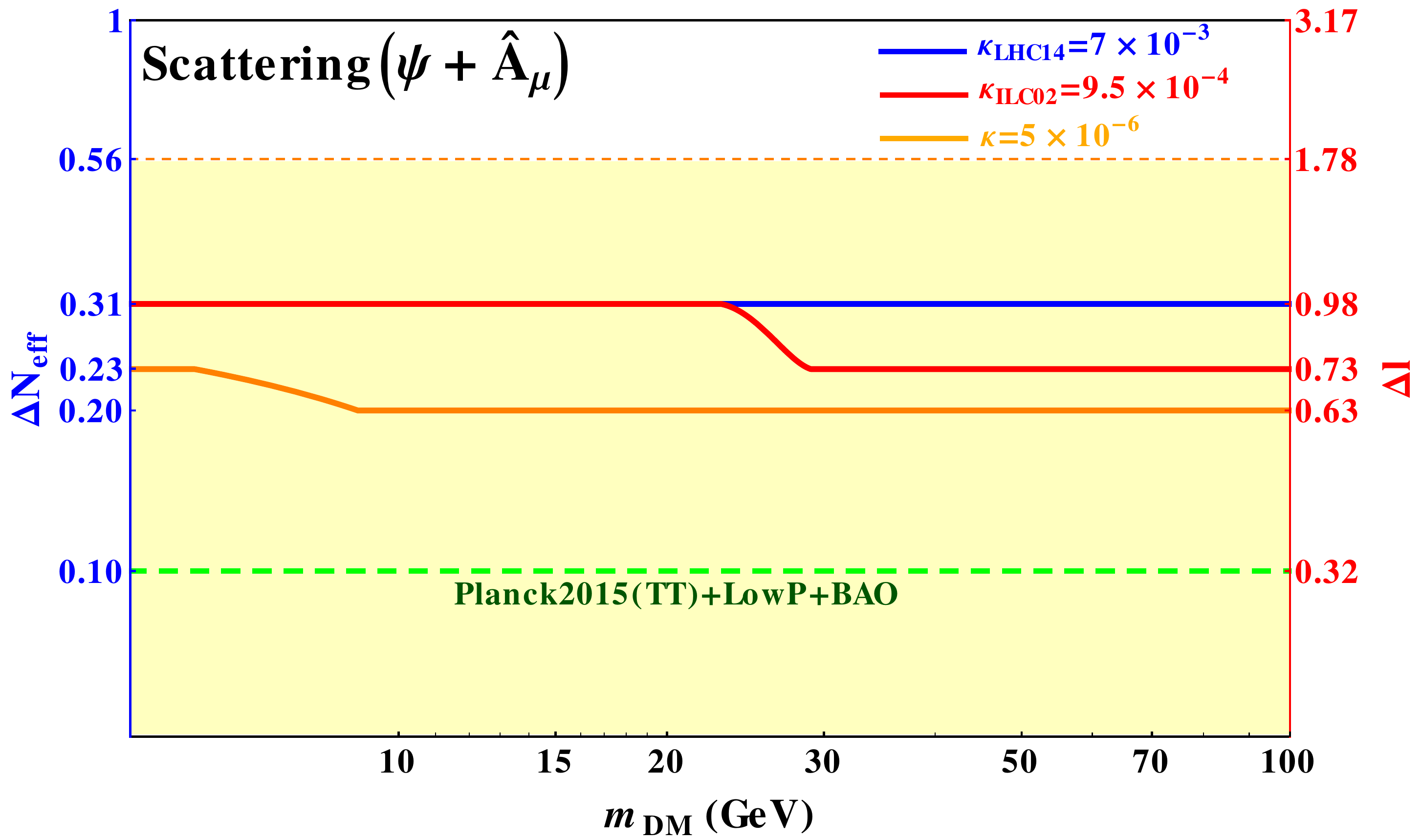}}
\mbox{\includegraphics[height=50mm]{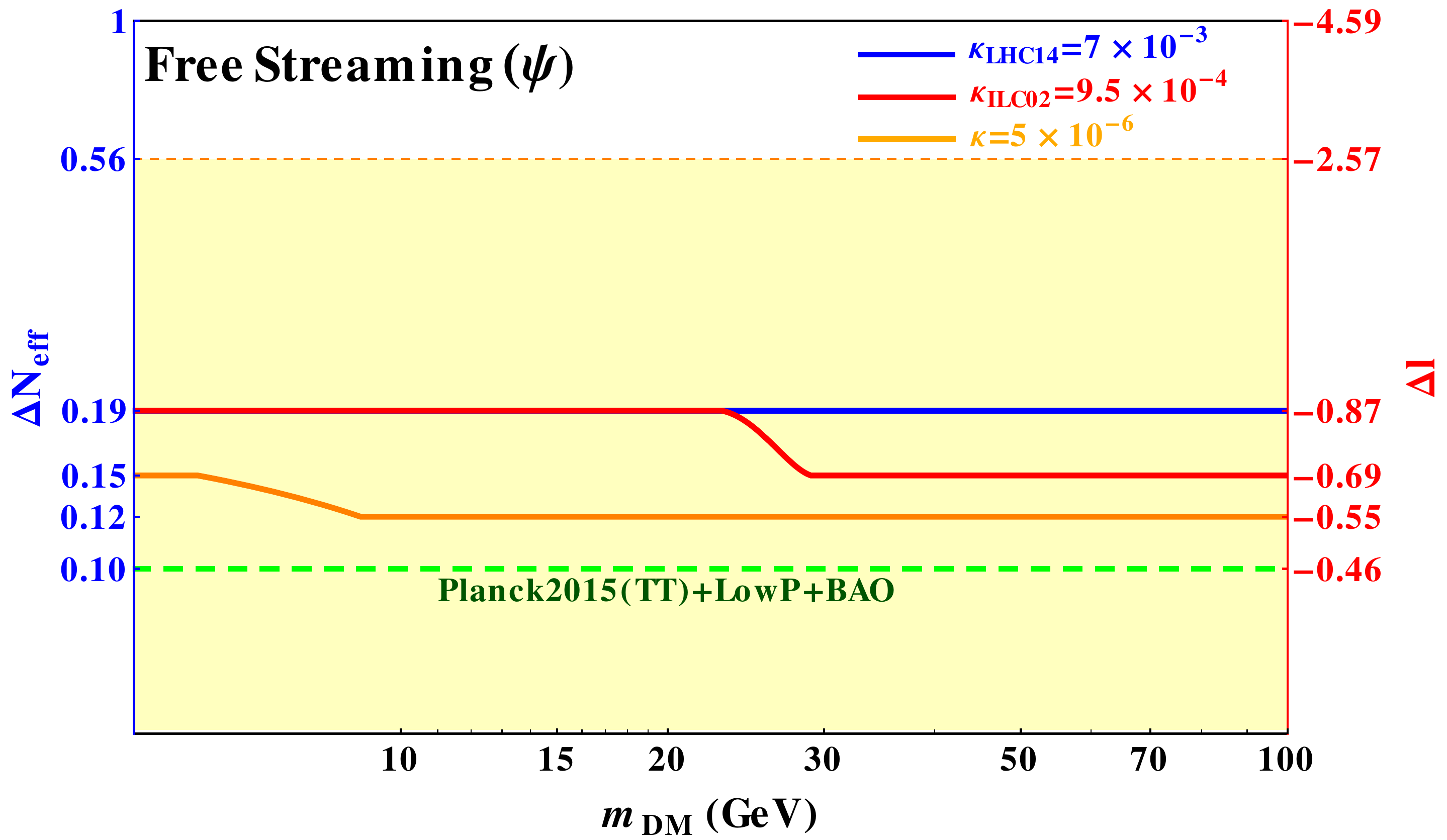}}
\caption{$\Delta N_{\text{eff}}$ and $\Delta \ell$ as a function of $m_{\text{DM}}$. It is assumed that $M_{\hat{Z}} > m_{\text{DM}}$. Three values of $\kappa$ are considered: $\kappa_\text{LHC14}, \kappa_\text{ILC02}$ and $\kappa = 5 \times 10^{-6}$. Also shown are the 2015 Planck results: the central value (Green dashed line) and the $2 \sigma$ constraint (Orange dashed line).} 
\label{fig:Observable_plot}
\end{figure}

Given the dark matter mass and the value of the coupling constant 
$\kappa$, we can determine the temperature $\Td$ at which the dark 
sector kinetically decouples from the SM\@. We can then use 
Eq.~(\ref{DeltaN}) to obtain $\Delta \Neff$ and Eq.~\eq{Delta_ell} to 
obtain the angular shifts of the CMB peaks. In our benchmark model, 
$\hat{g}_*$ and $\hat{g}_{*\text{kd}}$ are given by
 \begin{equation}
\hat{g}_* = 2 + \frac72 N_{\hat{\psi}}
\,,\quad
\hat{g}_{*\text{kd}} = n + \frac72 N_{\hat{\psi}}
\,.
 \end{equation}
 We assume that the $\hat{Z}$ boson is already nonrelativistic when the 
SM and hidden sectors decouple. Then $n=2$ if the DM 
candidate is already nonrelativistic when the SM and dark sectors 
decouple, and $n=4$ otherwise. 

The results for $N_{\hat{\psi} }= 1$ are shown in the upper panel of 
Fig.~\ref{fig:Observable_plot} for three different values of $\kappa$. The smallest of the 
values studied, $\kappa = 5 \times 10^{-6}$, corresponds to the minimum 
value of $\kappa$ that can ensure that the SM and hidden sector are in 
thermal equilibrium at or above the weak scale. The other two values 
studied correspond to the limits that can be placed on $\kappa$ at the 
LHC and at a lepton collider. In the figure we have plotted the current 
limits on $\Delta \Neff$ from the Planck experiment. Although in the 
region of light DM masses that may be probed at the 14 TeV LHC the 
predicted $\Delta \Neff$ are above the central value from Planck data 
fit, they are still well within $95\%$ C.L., and are large enough to be 
detected in upcoming experiments. It is also important to note that 
these bounds are not immediately applicable to this model since in the 
standard Planck analysis, it is assumed that the contribution to $\Delta 
\Neff$ is free streaming. A fresh analysis that relaxes this assumption 
by including scattering radiation is required to determine the current 
limits on this scenario.

We see from Fig.~\ref{fig:Observable_plot} that $\Delta \Neff \gtrsim 0.2$ in this class of 
models for the entire range of DM masses. Then, based on the discussion 
in Section II we expect that upcoming experiments will offer the 
possibility of distinguishing this scenario from one where $\Delta 
\Neff$ is the same, but the DR free streams. For the purposes of 
comparison, in the lower panel of Fig.~\ref{fig:Observable_plot} we have plotted $\Delta \Neff$ 
and $\Delta \ell$ for the same model, but without the U(1) gauge 
symmetry. The DR now consists of only one massless fermion species 
$\hat{\psi}$. Since the gauge boson $\hat{A}_\mu$ is now absent, the DR 
free streams rather than scatters. We see from the figure that although 
the magnitudes of $\Delta \Neff$ and $\Delta \ell$ are comparable in 
size to the benchmark model, the sign of $\Delta \ell$ is opposite in 
sign. It is this difference that we expect will help distinguish between 
free streaming and scattering DR\@.

\section{Conclusions}
\Secl{conclusion}

In this paper, we have explored a scenario where the DM candidate is 
part of a hidden sector whose particles carry no charges under the SM 
gauge groups, and whose couplings to the SM states, though present, are 
small. The abundance of DM is assumed to be determined primarily by 
annihilation to massless states that lie within the hidden sector. Then, 
if we further assume that the weak scale is the only mass scale in the 
hidden sector, so that the mass of the DM particle and its annihilation 
cross section are both set by this scale, the observed abundance of DM 
can be naturally explained. This framework is motivated by an 
alternative realization of thermal WIMP DM paradigm that is naturally 
compatible with limits from direct, indirect and collider searches for 
DM thus far.  

A robust consequence of this framework is the existence of DR, 
associated with the massless states in the hidden sector. The 
contribution of this DR to the energy density of the universe during the 
era of recombination epoch manifests itself observationally as a 
contribution to the effective number of neutrino species, $\Delta 
N_{\text{eff}}$. In addition, massless particles constituting DR may or 
may not interact with one another. We determined the shift in the 
locations of the CMB peaks, $\Delta \ell$, as a function of the number 
of free streaming DR species, $\Delta 
N_{\text{eff}}^{\text{free}}$, and scattering species, $\Delta 
N_{\text{eff}}^{\text{scatt}}$. We found that free streaming and 
scattering species shift the peaks in opposite directions, so that by 
combining this effect with the measurement of the total $\Delta 
N_{\text{eff}}$ mentioned above, we can separately determine $\Delta 
N_{\text{eff}}^{\text{free}}$ and $\Delta 
N_{\text{eff}}^{\text{scatt}}$. We also calculated the corrections to 
the amplitudes of the scalar and tensor modes of the CMB arising from the 
presence of DR, and showed that the sign of the correction depends on 
whether the DR scatters or free streams.

We have found that provided the hidden sector is initially in thermal 
equilibrium with the SM degrees of freedom at temperatures at or above 
the weak scale, there is a robust prediction for a lower bound on 
$\Delta N_{\text{eff}}$ of about 0.02. This is large enough to be 
observed by future CMB Stage-IV experiments. In the scenario where the 
DR has self-interactions, assuming naturalness, $\Delta N_{\text{eff}}$ 
is expected to lie well above this lower bound, making it large enough 
to be observed in upcoming experiments such as CMBPol. We further 
constructed a simple model that realizes this scenario. In our model, 
the SM and hidden sector are initially kept in thermal equilibrium 
through the Higgs portal. We have determined the size of the $\Delta 
N_{\text{eff}}$ signal, and shown that it is large enough to be detected 
in upcoming experiments. These experiments are also expected to be 
sensitive enough to the $\Delta \ell$ signal to allow the possibility of 
distinguishing between free streaming DR and scattering DR\@. In 
addition, we find that there are regions of parameter space where 
invisible decays of the Higgs into hidden sector states can be used to 
probe this model at colliders. 

An important point to reiterate is that the bounds on $\Delta 
N_{\text{eff}}$ derived from the standard analyses, such as by Planck, 
may not be directly applicable to our scenario where the DR may be 
self-scattering. A dedicated analysis parametrized by $\Delta 
N_{\text{eff}}^{\text{free}}$ and $\Delta N_{\text{eff}}^{\text{scatt}}$ 
instead of a single $\Delta N_{\text{eff}}$ is required to determine the 
exact limits on this scenario.

The general possibility that dark matter may arise from a sector hidden 
from the visible matter of the SM has drawn considerable interest 
recently, due in no small part to the increasing experimental limits on 
the conventional WIMP DM paradigm. Our work highlights that in this 
scenario which is challenging or perhaps even impossible for 
conventional DM searches, CMB experiments may be able to shed light on 
the nature of the DM sector, which is worthy of further exploration.

\acknowledgments
We thank Chris Brust for helpful discussions. ZC, YC and SH are supported by the NSF under grant PHY-1315155 and the 
Maryland Center for Fundamental Physics. YC is also supported by 
Perimeter Institute for Theoretical Physics, which is supported by the 
Government of Canada through Industry Canada and by the Province of 
Ontario through the Ministry of Research and Innovation. SH is also 
supported in part by a fellowship from The Kwanjeong Educational 
Foundation. TO is supported by the DOE under grant DE-FG02-13ER41942.

\vskip 1em
\noindent
{\bf Note added:} While we were completing this paper, we received~\cite{Buen-Abad:2015ova}, 
which overlaps with some of the ideas presented here.

%

\end{document}